\documentclass[12pt]{article}
\usepackage[utf8]{inputenc}
\usepackage[usenames,dvipsnames]{color}
\usepackage[normalem]{ulem}
\usepackage{amsmath}
\usepackage{amssymb}
\usepackage{mathptmx}
\usepackage{authblk}

\usepackage[a4paper, margin=1in]{geometry}

\usepackage[left]{lineno}

\allowdisplaybreaks

\usepackage{multirow}
\usepackage{array}
\usepackage{rotating}
\usepackage{booktabs}%
\usepackage{romannum}

\usepackage[breaklinks,colorlinks,citecolor=RoyalBlue,linkcolor=magenta]{hyperref}

\usepackage[super,compress,comma,sectionbib]{natbib}
\usepackage{graphicx}
\usepackage{subcaption}
\usepackage[labelfont=bf]{caption}
\DeclareCaptionLabelSeparator{bar}{ \textbar{} }
\def\jnl@style{\it}
\def\aaref@jnl#1{{\jnl@style#1}}

\newcommand{\apj}{Astrophys. J.}   
\newcommand{\apjl}{Astrophys. J. Lett.}   
\newcommand{\apjs}{Astrophys. J. Suppl. Ser.}   
\newcommand{\aap}{Astron. Astrophys.}   
\newcommand{\aapr}{Astron. Astrophys. Rev.}   
\newcommand{\mnras}{Mon. Not. R. Astron. Soc.}   
\newcommand{\pasp}{Publ. Astron. Soc. Pac.}   

\renewcommand{\figurename}{Fig.}

\newcommand{\msun}{\mbox{$M_\odot$}}

\newcommand{\mJybeam}  {\mbox{mJy}~\mbox{beam}^{-1}}

\newcommand{\kms}	{\mbox{km s}^{-1}}

\newcommand{\yr}	{{\rm yr}}

\newcommand{\au} {\mbox{au}}

\newcommand{\elecm}     {\mathrm{EM}}
\newcommand{\pccm}  {\mbox{pc cm}^{-6}}

\newcommand{\Jykms} {\mbox{Jy km s}^{-1}}

\newcommand{\kpc} {\mbox{kpc}}

\newcommand{\vlos}{v_\mathrm{los}}

\title{\textbf{
An eccentric massive protobinary assembled via a core merger parabolic encounter}}

\author[1,2,3]{Yao Wang}

\affil[1]{\small State Key Laboratory of Dark Matter Physics, School of Physics and Astronomy, Shanghai Jiao Tong University, Shanghai 200240, China}

\affil[2]{\small Department of Astronomy, School of Physics and Astronomy, Shanghai Jiao Tong University, 800 Dongchuan Road, Shanghai 200240, China}

\affil[3]{\small Key Laboratory for Particle Astrophysics and Cosmology (MOE) / Shanghai Key Laboratory for Particle Physics and Cosmology, Shanghai 200240, China}

\author[1,2,3*]{Yichen Zhang}

\author[4]{Rub{\'e}n Fedriani}
\affil[4]{\small Instituto de Astrof\'isica de Andaluc\'ia, CSIC, Glorieta de la Astronom\'ia s/n, 18008 Granada, Spain}

\author[5]{Kei E. I. Tanaka}
\affil[5]{\small Department of Earth and Planetary Sciences, Institute of Science Tokyo, Meguro, Tokyo, 152-8551, Japan}

\author[6,7]{Viviana Rosero}
\affil[6]{\small Cahill Center for Astronomy and Astrophysics, MC 249-17, California Institute of Technology, Pasadena, CA 91125, USA}
\affil[7]{\small Space Science Institute, 4750 Walnut Street, Suite 205, Boulder, CO 80301, USA}

\author[1,2,3]{Kai Yang}

\author[8]{Morten Andersen}
\affil[8]{European Southern Observatory, Karl Schwarzschild Str. 2, 85748 Garching, Germany}

\author[9]{Maria T. Beltr\'an}
\affil[9]{\small INAF-Osservatorio Astrofisico di Arcetri, Largo E. Fermi 5, I-50125 Firenze, Italy}

\author[10]{M\'elisse Bonfand}
\affil[10]{\small Department of Astronomy, University of Virginia, Charlottesville, Virginia 22904, USA}

\author[11]{Yu Cheng}
\affil[11]{\small National Astronomical Observatory of Japan, 2-21-1 Osawa, Mitaka, Tokyo, 181-8588, Japan}

\author[12]{James M. De Buizer}
\affil[12]{\small Carl Sagan Center for Research, SETI Institute, Mountain View, CA, USA}

\author[1,2,3]{Yihuan Di}

\author[13,14]{Guido~Garay}
\affil[13]{\small Departamento de Astronom\'ia, Universidad de Chile, Camino el Observatorio 1515, Las Condes, Santiago, Chile}
\affil[14]{\small Chinese Academy of Sciences South America Center for Astronomy, National Astronomical Observatories, Chinese Academy of Sciences, Beijing, 100101, China}

\author[15,16,17]{Prasanta Gorai}
\affil[15]{\small Universit\"at Heidelberg, Zentrum f\"ur Astronomie, Institut f\"ur Theoretische Astrophysik, Albert-Ueberle-Str. 2, 69120 Heidelberg, Germany}
\affil[16]{\small Rosseland Centre for Solar Physics, University of Oslo, PO Box 1029 Blindern, 0315, Oslo, Norway}
\affil[17]{\small Institute of Theoretical Astrophysics, University of Oslo, PO Box 1029 Blindern, 0315, Oslo, Norway}

\author[10]{Zhi-Yun Li}

\author[18]{Yao-Lun Yang}
\affil[18]{\small Star and Planet Formation Laboratory, RIKEN Pioneering Research Institute, Wako-shi, Saitama, 351-0198, Japan}

\author[10,19]{Jonathan~C.~Tan}
\affil[19]{\small Department of Physics and Astronomy, Chalmers University of Technology, 412 93 Gothenburg, Sweden}

\affil[*]{\small e-mail: \url{yczhang.astro@gmail.com}}

\date{}
 
\begin{document}

\pagenumbering{arabic}

\maketitle

{\color{blue}
\noindent
Most massive stars form in binary systems, which profoundly influence their subsequent evolution. However, how such systems form remains poorly understood, with several competing scenarios proposed, including disk fragmentation, core fragmentation and capture. 
Determining the orbital architectures of massive binaries, particularly during their earliest embedded phases, is therefore crucial for distinguishing among these formation pathways, but direct measurements of their three-dimensional motions have remained exceptionally challenging. 
Here we present high-resolution, multi-epoch sub-millimeter-to-centimeter ALMA and JVLA observations of the massive protobinary IRAS 07299$-$1651, complemented by JWST and VLT infrared imaging. 
We detect orbital proper motion of the binary components, enabling a full three-dimensional orbital reconstruction. 
Combining orbital fitting, multi-wavelength continuum modelling, hydrogen recombination line kinematics and jet observations, we find that the preferred orbital solutions are highly eccentric and close to parabolic, while both circumstellar disks are strongly misaligned with the orbital plane. 
These properties are naturally explained by a ``core-merger'' scenario in which the two protostars originated independently from initially unbound cores that recently underwent a near-parabolic encounter, producing an eccentric binary with a current separation of about 200 au. These findings suggest that the core-merger process may represent an important pathway for forming eccentric massive binaries.

}

It is estimated that at least 90\% of massive stars exist in binary or higher-order 
multiple systems \cite{Chini12,Peter12,Sana14,Moe17}, 
and this fraction is even higher for newly formed massive stars \cite{Pomohaci19,Bordier22}. 
Understanding the formation of massive binaries is therefore a fundamental problem in star formation\cite{Kratter10}.
Observational studies have largely focused on relatively evolved systems, 
when the natal cores have largely dispersed and the protostars are exposed \cite{Bordier22,kraus_high-mass_2017,Koumpia21},
whereas closely separated embedded massive protobinaries remain rare 
owing to observational challenges, 
despite recent discoveries at millimeter and radio wavelengths
\cite{beltran_binary_2016,beuther_multiplicity_2017,zhang_dynamics_2019,zapata_asymmetric_2019,Cyganowski22}.
IRAS 07299$-$1651 is one of these rare, early-stage massive protobinaries. 
Located at $1.68\pm0.1~\kpc$\cite{distance1680}, it is surrounded and fed by several large-scale stream-like structures 
and exhibits an apparent separation of $\sim$180~au. 
First identified in 1.3~mm continuum observations 
by the Atacama Large Millimeter/submillimeter Array (ALMA)\cite{zhang_dynamics_2019}, 
the system shows a line-of-sight velocity difference of $\sim 10~\kms$ 
between its two components in the H30$\alpha$ hydrogen recombination line (HRL), 
providing a rare dynamical constraint on the orbit of an embedded massive protobinary.

\vspace{1em}
\noindent{\textbf{Discovery of orbital proper motion}}

We obtained high-resolution ($\sim 0.05'',~84~\mathrm{au}$) ALMA ($0.3-3$~mm) and 
the Karl G. Jansky Very Large Array (JVLA) (7 and 13~mm) continuum observations of 
IRAS 07299$-$1651 over 2016$-$2024 (\hyperref[sec:methods]{Methods}). 
Both components are spatially resolved 
(Fig.~1a; Extended Data Fig.~1); 
the brighter northwestern source is referred to as Source A and the other as Source B.
Assuming the 0.3~mm emission is dominated by dust continuum with a spectral index of 2.5, 
we measured dust-subtracted continuum spectral indices. 
Source A shows $\alpha_\mathrm{A}=1.55\pm0.17$ at lower frequencies 
and $\alpha_\mathrm{A}=0.57\pm0.08$ at higher frequencies, 
while Source B has $\alpha_\mathrm{B}=1.18\pm0.04$ (Extended Data Fig.~2),
consistent with partially optically thick free-free emission from ionized gas. 
Hydrogen recombination lines 
(H42$\alpha$, H40$\alpha$, H30$\alpha$, H26$\alpha$, H19$\alpha$) 
are also detected toward both sources 
(Fig.~1b,c; Extended Data Fig.~3).
These free-free and HRL emissions originate from compact ionized regions 
very close to each forming star; they not only trace
the circumstellar structures, but also the stellar motions themselves.

By two-component Gaussian fitting to the continuum images
(\hyperref[sec:methods]{Methods}; Supplementary Fig.~1; Supplementary Table~1), 
we determined the positions of the two sources with a typical accuracy of $\sim0.1~\text{mas}$.
Source B is moving away from Source A in the plane of the sky (Fig.~1d). 
Linear fits to the relative position changes over time (Fig.~1e,f) yield 
a total relative proper-motion velocity of $10.4\pm1.6~\kms$ ($1.3\pm0.2~\mathrm{mas~yr^{-1}}$). 
To account for systematic uncertainties beyond those estimated from image fitting, 
we include an isotropic intrinsic-scatter term, $\sigma_{\rm int}$, in the fit. 
The resulting value, $\sigma_{\rm int}=1.3^{+0.5}_{-0.3}$ mas, corresponds to a reduced $\chi^2$ of 1.1.
Furthermore, visibility-domain fitting (\hyperref[sec:methods]{Methods}; Supplementary Figs.~2 and 3) 
provides positions consistent with the image-based results 
within the inferred intrinsic scatter.
The reliability of the position determination is further discussed in Discussion 1 in Supplementary Information.

Gaussian fitting of the HRL spectra 
(\hyperref[sec:methods]{Methods}; Extended Data Fig.~3; Supplementary Table~2) 
shows that the H26$\alpha$ central velocities are highly consistent with 
previous H30$\alpha$ measurements \cite{zhang_dynamics_2019} 
($\sim 15~\kms$ for Source A and $\sim 25~\kms$ for Source B), 
confirming a $\sim10~\kms$ line-of-sight velocity difference. 
The relative proper motion velocity of $10.4\pm1.6~\kms$ 
is comparable to the radial velocity difference between the two sources, 
consistent with the scenario of orbital motion.
Alternative explanations for the observed positional shifts, 
including wavelength-dependent centroid displacements and jet variability, 
are discussed in Discussion 2 in Supplementary Information and are unlikely to account for the observed motion.


\vspace{1em}
\noindent{\textbf{Ionized circumstellar disks}}

From the Gaussian fitting to the continuum images, 
we also obtained the deconvolved shapes of the two sources at each wavelength,
which are also confirmed by visibility fitting (Supplementary Fig.~3). 
In both sources, the deconvolved properties -- including size, position angle, and aspect ratio -- 
remain remarkably consistent from 3~mm to~0.6 mm, 
despite the varying shapes of the synthetic beams 
(Fig.~1a; Extended Data Fig.~4; also see Discussion 1 in Supplementary Information). 
This stability indicates that the deconvolved ellipses reflect the intrinsic source morphologies 
at these wavelengths. 
Assuming the ellipses correspond to inclined disks, 
we estimated that the two disks have major axis position angles of 
$18^{\circ}\pm4^{\circ}$ (Source A) and $68^{\circ}\pm8^{\circ}$ (Source B),
and inclinations with respect to the plane of sky of 
$i=29^{\circ}\pm5^{\circ}$ for Source A and $i=61^{\circ}\pm2^{\circ}$ for Source B.
Adopting the $2\sigma$ deconvolved semi-major axis as a proxy for the disk radius\cite{tobin2020}, 
both disks show comparable outer radii of $\sim 20~\au$, 
providing an estimate of the characteristic scale of the ionized disks. 
A more extended neutral dusty disk is expected, as indicated by a larger size of $\sim 40~\au$
inferred from the 0.3 mm data using the same approach.

The HRL kinematics provide independent confirmation of ionized circumstellar disks. 
In both sources, the HRL centroid distributions are highly ordered
and symmetric with respect to the continuum peak and the minor axes 
of the deconvolved continuum from 3 to 0.6~mm (Fig.~1b,c), 
consistent with disk rotation on $\sim 10~\au$ scales. 
Fitting simple Keplerian rotation models \cite{cesaroni_dissecting_2025} to these centroids 
(\hyperref[sec:methods]{Methods}; Fig.~2) 
confirms the rotation pattern and 
yields dynamical central masses of $M_\mathrm{A}=12.1^{+4.9}_{-2.9}~M_\odot$ and $M_\mathrm{B}=12.1^{+5.5}_{-5.0}~M_\odot$. 
The actual masses may be somewhat higher if the disks are sub-Keplerian. 
The HRL centroid fits also independently validate the disk position angles and 
inclinations inferred from the continuum deconvolution.

\vspace{1em}
\noindent{\textbf{Jet and disk orientations}}

The presence of circumstellar disks is further supported by jets detected at radio and infrared wavelengths. A radio jet is seen in low-resolution 1.3 and 6~cm continuum 
(Extended Data Fig.~5). 
While the resolution is insufficient to directly resolve the driving sources, 
the jet position angle ($\sim -72^\circ$) is perpendicular to the Source A disk major axis 
($\mathrm{P.A.}\approx18^\circ$), strongly indicating that Source A drives the radio jet. 
No secondary jet is detected in the radio data.

At larger scales, the primary outflow is traced by H$_2$ 1$-$0 S(1) ($2.12~\mu$m) and 
0$-$0 S(9) ($4.69~\mu$m) emission in the James Webb Space Telescope (JWST) images 
(\hyperref[sec:methods]{Methods}; Fig.~3a,b). 
A chain of H$_2$ knots extends westward, culminating in a bow-shaped feature 
($\mathrm{P.A.}\approx -66^\circ$; Fig.~3c), 
consistent with the radio jet and Source A disk orientation. 
However, due to the large and strong nebulous emission, 
the detection of the inner knots is hampered by the over-subtraction in the difference image 
(Fig.~3c). 
The inner knots of the outflows are better recovered in VLT K-band Multi Object Spectrograph (KMOS) integral-field-unit (IFU) data 
(\hyperref[sec:methods]{Methods}; Fig.~3d,e), 
revealing a bright H$_2$ knot coinciding with the radio jet knot to the west (Fig.~3e). 
In addition, southeast of the sources, one knot at $\mathrm{P.A.}\sim166^\circ$ 
(Fig.~3e) is nearly 
perpendicular to the Source B disk major axis ($\mathrm{P.A.}\approx68^\circ$), 
possibly tracing a secondary jet from Source B. 
These radio and IR observations confirm the presence of two circumstellar disks,
each driving its own jet, and
support the disk orientations inferred from continuum morphology and HRL kinematics.

These jet observations not only support the presence of circumstellar disks, 
but also provide information on their three-dimensional orientations.
While the disk inclination angles are relatively well constrained, their signs remain ambiguous.
Assuming that the brighter lobe preferentially traces the near-facing side of a bipolar jet\cite{fedriani2018}, 
the northwestern side of the Source A jet and the southeastern side of the Source B jet are inferred to be tilted toward the observer.
For Source A, this interpretation is independently supported by the C-shaped distribution of HRL centroids 
toward the southeastern half of the disk (Fig.~2), 
likely caused by optical depth variations enhancing emission from the near half of the disk 
\cite{cesaroni_dissecting_2025}. 
In this interpretation, the southeastern half of the Source~A disk is closer to the observer, 
consistent with that the northwestern jet being the near-facing side.
These constraints enable the three-dimensional orientations of the two circumstellar disks to be determined.
However, the observed brightness asymmetry of the jet lobes may be influenced by other factors 
such as the ambient gas distribution. 
The sign of the inclination angle therefore remains only weakly constrained.

\vspace{1em}
\noindent{\textbf{SED and HRL flux fitting}}

To further constrain the nature of the two forming stars, 
we performed a joint fit to the continuum spectral energy distributions (SEDs) 
and multiple HRL fluxes for both sources (\hyperref[sec:methods]{Methods}). 
The model consists of inclined ionized circumstellar disks 
with an additional dust emission component, 
treating continuum and HRL optical depths self-consistently and 
assuming local thermodynamic equilibrium (LTE) for the HRLs (see Discussion 3 in Supplementary Information).
The fits are shown in Fig.~4, 
while the best-fitting parameters 
and their uncertainties
are listed in Extended Data Table~1 (see also Supplementary Figs. 4 and 5).

The inferred outer radii of the ionized disks are 
$33^{+26}_{-7}$ au and $22^{+24}_{-6}$ au for Sources A and B, respectively, 
consistent with the characteristic radii of $\sim20$ au derived from the continuum images. 
The disk inclinations are less tightly constrained by the SED and HRL fitting, 
but are consistent with those inferred from the continuum morphology.

The best-fit models imply ionizing photon rates of 
$3.6^{+1.8}_{-1.2}\times10^{46}$ s$^{-1}$ and 
$1.6^{+0.8}_{-0.5}\times10^{46}$ s$^{-1}$ 
for Sources A and B, respectively. 
If interpreted using zero-age main-sequence (ZAMS) models\cite{davies_red_2011}, 
these correspond to stellar masses of 
$13.4\pm0.5~M_\odot$ and $12.5^{+0.5}_{-0.4}~M_\odot$,
and spectral types between B1 and B0\cite{mottram_rms_2011}.
Combining the observed bolometric luminosity of $6.5^{+6.6}_{-3.3}\times10^4~L_{\odot}$\cite{Fedriani2023}
and the derived ionizing photon rates with 
protostellar evolutionary models spanning a range of accretion histories\cite{zhang2014,tanaka_outflow-confined_2016,zhang_radiation_2018},
we constrain the total mass to be $17.7-28.1~M_{\odot}$. 
The corresponding protostellar mass range is $8.5 - 16.3~M_{\odot}$ for Source A 
and $7.7 - 16.5~M_{\odot}$ for Source B
(the two ranges cannot be directly added, because of correlated uncertainties).
These ranges are in good agreement with the independent dynamical masses inferred from the HRL kinematics.


\vspace{1em}
\noindent{\textbf{Orbital architecture and disk-orbit misalignment}}

Using the measured orbital proper motions and radial velocities, 
we performed a unified binary orbit fit allowing elliptical, parabolic, and hyperbolic solutions 
(Orbital Fitting 1; \hyperref[sec:methods]{Methods}; Extended Data Table~2).
The inferred eccentricity shows a clear dependence on the total system mass:
lower masses preferentially yield hyperbolic solutions, whereas higher masses favor bound orbits
(Fig.~5a; Supplementary Fig.~6).

As discussed above, independent constraints converge on a consistent system mass. 
HRL kinematics yield a total dynamical mass of $24.9^{+7.5}_{-6.3}~M_{\odot}$.
The disk SED+HRL fitting gives a ZAMS-based mass of $25.9\pm0.7~M_{\odot}$, 
while protostellar evolutionary models imply a total mass range of $17.7 - 28.1~M_{\odot}$. 
In addition, infall kinematics of the surrounding large-scale gas provide 
another independent estimate of $27\pm6~M_{\odot}$\cite{zhang_dynamics_2019}. 
Within this overlapping mass range, and particularly near $M_{\mathrm{tot}}\approx25~M_{\odot}$,
the eccentricity distributions peak close to unity (Fig.~5a),
indicating that the preferred solutions are near-parabolic.

To further test this result, we performed unified orbit fit for a constrained mass range 
(Orbital Fitting 2; Supplementary Fig.~7)
using a Gaussian prior for $M_\text{tot}$ based on the SED+HRL fitting ($M_{\rm tot}=25.9\pm4.0~M_\odot$),
bounded by the allowed evolutionary-model range of $17.7$--$28.1~M_\odot$.
The resulting eccentricity posterior again peaks near unity 
(Extended Data Fig.~6a), 
further supporting a preference for near-parabolic solutions.
The inferred intrinsic-scatter term remains consistent with that obtained from the proper-motion fitting.

Despite these constraints, the eccentricity remains only loosely constrained 
owing to the limited orbital coverage.
First, the inferred eccentricity depends sensitively on the system mass 
(Fig.~5a): bound solutions are favored if the total mass reaches $\gtrsim30~M_\odot$. 
It is therefore important to assess whether the independently inferred masses could be substantially underestimated.
As discussed in ``Ionized circumstellar disks'', the masses derived from HRL kinematics may be underestimated.
Furthermore, if dust attenuation within the ionized regions is substantial, 
or if large portions of the disks remain optically thick at all observed wavelengths, 
the ionizing photon rates -- and thus the stellar masses inferred from the combined SED and HRL fitting --
would also represent lower limits. 
However, owing to the steep dependence of ionizing photon rate and bolometric luminosity on (proto)stellar mass, 
the true masses are unlikely to exceed the inferred range by a large margin.
Second, even within the independently inferred mass range of $\sim20$--$30~M_\odot$, 
lower-eccentricity bound solutions remain viable (Discussion 4 in Supplementary Information). 
Therefore, although near-parabolic solutions are preferred by the current data,
continued astrometric monitoring will be required to further refine the orbital parameters
and establish the orbital eccentricity more definitively.
We also performed orbital fitting using only the radial velocity and single-epoch astrometric constraints 
available in the previous study of this source \cite{zhang_dynamics_2019}
(Orbital Fitting 3, Discussion 4 in Supplementary Information).
Compared with that fit, the newly measured proper motion provides a critical additional constraint 
on the orbital architecture and substantially strengthens the preference for high-eccentricity solutions.

To illustrate the preferred class of solutions,
Fig.~5(b--d) shows representative parabolic orbits obtained under the constrained mass range 
(Orbital Fitting 4; Supplementary Fig.~10).
The fit yields a periapsis distance of $q = 139^{+88}_{-62}$~au and a periapsis passage time of $t_p = 39^{+22}_{-7}\times10^3$~MJD.
The inferred periapsis occurred $58^{+18}_{-61}$ years before the observational epochs (2016--2024), indicating that the system is currently observed close to its point of closest approach. The true separation of the binary at the observational epochs is $s_{\rm 3D}\approx220$ au.
The maximum-likelihood parabolic orbit (black curve) has $q=122$ au and a periapsis passage approximately 60 years ago, although the exact maximum-likelihood solution is sensitive to posterior sampling.
The fitted parameters are listed in Extended Data Table~3.


Combining the measured position angles and inclinations of the two circumstellar disks, 
we constrain their three-dimensional orientations relative to the orbital plane
(\hyperref[sec:methods]{Methods}; Fig.~6a,b). 
For generality, we adopt the posterior samples from the unified orbital fit with a constrained mass range
(Fitting 2) to infer the orbital orientation.
The angle between the Source A disk and the orbital plane is
$\psi_\mathrm{A,orb}=\left(59^{+17}_{-8}\right)^\circ$, 
while that for Source B
$\psi_\mathrm{B,orb}=\left(133^{+23}_{-18}\right)^\circ$, 
where angles $>90^\circ$ indicate retrograde rotation with respect to the orbital motion. 
The angular momentum vectors of the two disks are themselves strongly misaligned, 
with $\psi_\mathrm{A,B}=99^\circ\pm5^\circ$.
The quoted values assume the inclination signs inferred from the infrared jet asymmetries,
i.e. the southeastern half of the Source A disk 
and the northwestern half of the Source B disk are closer to the observer.
Alternative configurations are shown in Extended Data Fig.~7; 
the conclusion that the system is strongly misaligned is robust in all cases.

\vspace{1em}
\noindent{\textbf{Implications for massive binary formation}}

Three main pathways have been proposed for binary formation \cite{Offner22}: 
disk fragmentation, in which a protobinary forms through gravitational instability 
in an accretion disk \cite{Kratter06,Krumholz09}; 
core or filament fragmentation, in which each component originates from independent fragmentation 
within a bound parent cloud \cite{Bate12}; 
and capture, in which initially unbound stars become gravitationally bound through interactions 
with surrounding stars or gas \cite{Munoz15}. 
Previous observational and theoretical studies have suggested that disk fragmentation 
preferentially produces close binaries with separations $\lesssim500$ au 
and approximately aligned angular momentum vectors, 
whereas core fragmentation more commonly yields wider and misaligned systems
\cite{Offner22,Offner1O,Tobin16,Oliva20}.

Despite its relatively small present-day separation, the preferred orbital solutions for IRAS 07299$-$1651 are 
close to parabolic or highly eccentric, 
and the system exhibits large misalignments between the binary orbital plane and both circumstellar disks.
Moreover, no circumbinary disk is detected in this system.
These properties are difficult to reconcile with a disk fragmentation origin. 
Even when restricted to elliptical solutions, the fitted orbits remain highly eccentric and 
strongly misaligned, in tension with expectations from disk fragmentation. 
Recent simulations incorporating highly dynamical, turbulent accretion flows have shown that 
disk fragmentation can occasionally produce compact, misaligned binaries \cite{tu_fragmentation_2024}; 
however, the near-parabolic or extremely eccentric orbital solutions in IRAS 07299$-$1651 disfavor this scenario.

Instead, the preferred orbital solutions place the system close to the boundary between bound and unbound configurations. 
These properties are consistent with a scenario in which the two stars formed independently at much larger separations, 
either through fragmentation within marginally bound regions of a common cloud or from initially unbound cores,
before undergoing a recent close encounter. However, the core or filament fragmentation scenario requires synchronization of the massive star formation events within a relatively short period, which a priori makes this case appear to be unlikely. On the other hand, mergers of independent cores in a star-forming clump are expected to be a relatively common occurrence. For example, in the Turbulent Core Accretion model\cite{mckee03} a massive protostellar core is estimated to interact with an approximately equal mass of surrounding clump gas during the timescale of its collapse and such interaction is likely to be in the form of core mergers.

In the representative parabolic solutions (Orbital Fitting 4), 
the inferred separation $\sim10^5$ years ago 
-- comparable to the timescale of massive star formation -- 
was $(3.5\pm0.1)\times10^4$ au ($0.17\pm0.01$ pc). 
This scale is consistent with the formation of the two protostars in separate cores 
followed by a subsequent close interaction. 
In this picture, the present-day system may represent the outcome of a ``core-merger'' capture event.
For hyperbolic solutions in Fitting 2, 
the most probable relative velocity at infinity is $\lesssim 8~\kms$ 
(Extended Data Fig.~6b). 
This is comparable to the relative velocities expected between independently forming cores in turbulent massive clumps, 
where observed velocity dispersions of $\sigma_{\mathrm{1D}} \sim 1-2~\kms$ \cite{mckee03,Plume97}
imply characteristic three-dimensional relative velocities of $\sim2.5$--$5~\kms$.
This is comparable to the initial relative velocities 
inferred from the hyperbolic solutions.
Alternatively, for elliptical solutions in Fitting 2, 
highly eccentric orbits with long periods of
$\gtrsim 10^{3.5}-10^5~\yr$ are favored 
(Extended Data Fig.~6b).
These orbits may have already lost orbital energy through 
interactions with the surrounding gas during the merging event.
The pronounced misalignments between the orbital plane and the two circumstellar disks arise naturally in the ``core merger'' picture. 
Additional observational evidence is broadly consistent with this interpretation.
At least two large-scale stream-like structures are observed\cite{zhang_dynamics_2019}
toward the central sources (see Fig.~6c), 
with the northern stream carrying angular momentum broadly consistent with that of the Source A disk, 
plausibly tracing remnants of the larger-scale structures from which the two protostars assembled.


If such a close encounter occurred, an important question is whether the observed circumstellar disks 
could have survived it.
Various theoretical studies have shown that parabolic encounters can alter 
the inclinations and sizes of circumstellar disks, 
and induce substructures, 
depending on the periapsis distance, stellar mass ratio, 
and the relative orientation between disks and the orbit
\cite{flybyinclination2016,flybytruncation1993,Bhandare16, flybytruncation17,flybyspiral2019}.
However, for near-equal-mass encounters, the impact on disk--orbit alignment 
is expected to be modest \cite{flybyinclination2016}. 
We therefore interpret the observed disk-orbit misalignments as primarily primordial rather than induced by the recent encounter. 
This interpretation is supported by the close alignment between the Source A jet and disk. 
The reconstructed trajectories imply only modest positional changes over the past $\sim4\times10^3$ yr, 
comparable to the dynamical age of the jet (Fig.~6c), 
suggesting that the disk orientation has not been substantially altered during this period.

The observed disk properties are broadly consistent with this picture.
Adopting the periapsis-distance posterior of $139^{+88}_{-62}$ au from the fiducial fit (Orbital Fitting 4), 
theoretical calculations predict truncation radii of $54^{+24}_{-19}$ au \cite{Bhandare16}, 
consistent with the observed radii of $\sim20$ au for the ionized disks and $\sim40$ au for the dusty disks. 
By contrast, tidal truncation in a periodically interacting bound binary\cite{Truncation2005,Truncationusage2018}
would predict smaller characteristic disk sizes ($R_{\rm t}=32^{+12}_{-14}$ au) for the elliptical solutions from Fitting 2, 
favoring a one-time encounter over a long-lived bound configuration. 
In the most disruptive case of a prograde, coplanar, equal-mass encounter,
the disk rotation pattern is expected to remain intact within roughly 
one-quarter of the periapsis distance\cite{flybytruncation17}. 
The observed smooth disk structure with a well-defined rotational pattern 
on scales of $\sim20~\au$ is broadly consistent with this expectation,
although finer substructures or deviations from simple rotation 
cannot be resolved with the current data.

The future evolution of the system remains uncertain. 
Rapid orbital decay driven by disk interactions is unlikely 
because the two disks are well separated relative to their sizes \cite{Munoz15}. 
Capture through N-body interactions is also improbable given the system's isolation. 
Although the sources are currently moving apart, 
their relative motion places the system close to the boundary between bound and unbound states. 
Continued interaction with large-scale inflows may therefore further modify the orbital properties 
and potentially drive the system toward a more tightly bound state.

So far, only a small number of massive protobinaries with separations $\lesssim1000$ au 
have been identified \cite{kraus_high-mass_2017,beltran_binary_2016,beuther_multiplicity_2017,zhang_dynamics_2019,zapata_asymmetric_2019,sridharan_direct_2005,tanaka_salt_2020}, 
and orbital proper motions have so far been measured primarily in low-mass systems \cite{hernandezgarnica_accurate_2024,maureira_orbital_2020},
with only tentative evidence reported in massive forming binaries\cite{kraus_high-mass_2017}.
Misaligned circumstellar disks are well documented in low-mass protobinaries \cite{jensen_misaligned_2014,brinch_misaligned_2016,takakuwa_spiral_2017,ichikawa_misaligned_2021,diaz-rodriguez_physical_2022}, but remain
poorly constrained observationally in the massive regime\cite{kraus_high-mass_2017,beuther_multiplicity_2017}.
IRAS 07299$-$1651 therefore provides a rare system in which the three-dimensional orbital architecture 
and the orientations of both circumstellar disks can be constrained simultaneously using multiple independent diagnostics. 
The system is deeply embedded and heavily obscured at near- and mid-infrared wavelengths, 
indicating an early evolutionary stage. 
These characteristics make IRAS 07299$-$1651 a valuable laboratory for investigating the formation of hundred-au-scale massive binaries 
and the possible role of core-merger capture events in massive binary formation.

\vspace{1em}

\begin{figure}
\centering
\includegraphics[width=0.85\textwidth]{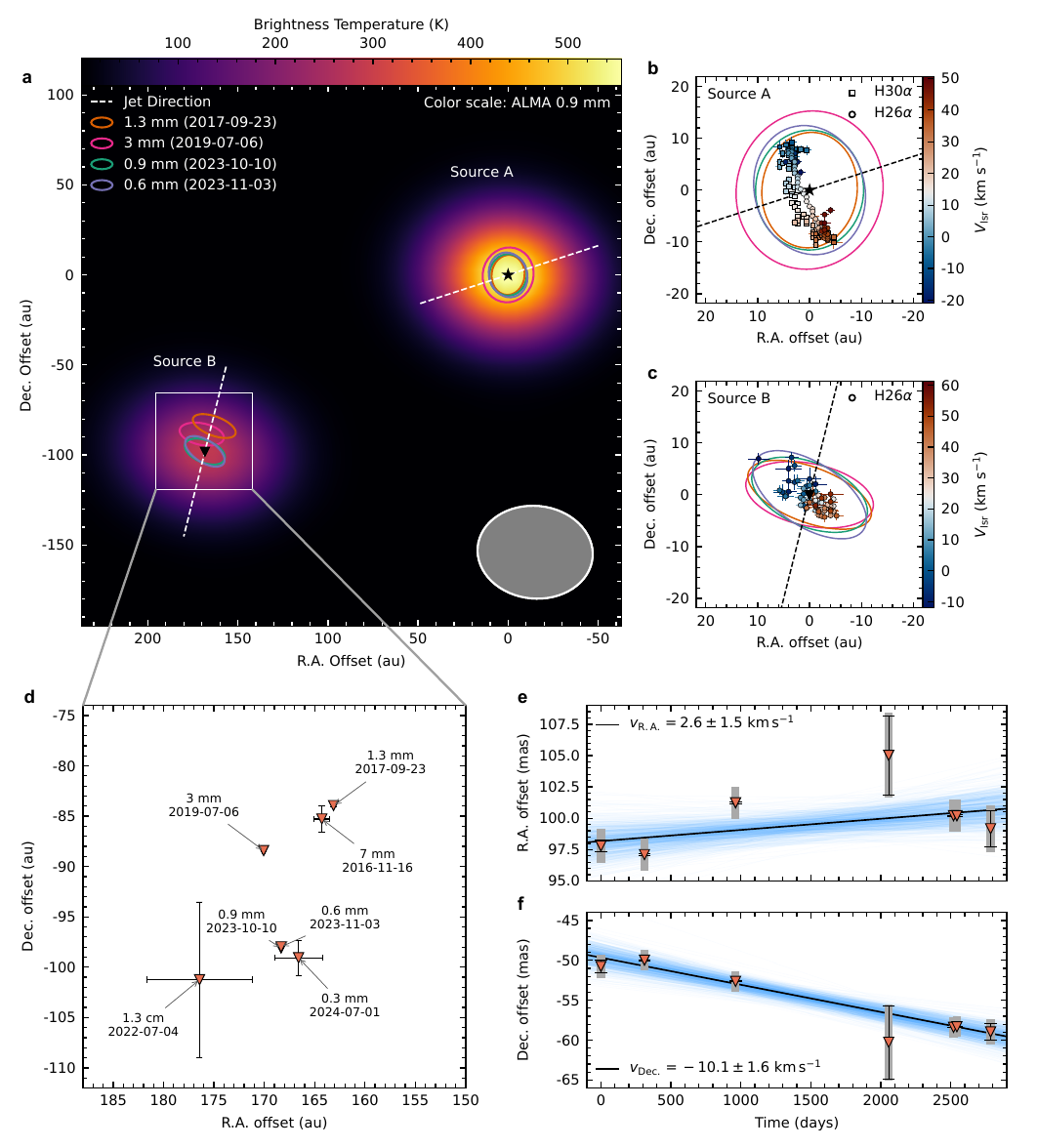}
\caption{
\textbf{Multi-wavelength continuum emission, hydrogen recombination line kinematics and relative astrometry of the protobinary system.}
{\bf a}, ALMA 0.9~mm continuum (background) overlaid with 
deconvolved 2D Gaussian FWHM profiles (ellipses) from 3 to 0.6~mm for both sources. 
Positions of Source A at different wavelengths are aligned and marked with a star. 
Its absolute position at 0.9 mm is 
$\left(\alpha_{\rm ICRS},\delta_{\rm ICRS}\right)_{\rm A} = (07^{\rm h}32^{\rm m}09.783^{\rm s}, -16^\circ58^\prime12^{\prime\prime}.12)$. Here ICRS denotes the International Celestial Reference System.
Inverted triangles mark the positional offsets of Source B relative to Source A; physical scales assume a distance of $1.68~\kpc$ (also used in panels b--d).
The grey ellipse at the lower-right corner shows the 0.9~mm synthesized beam 
(see Extended Data Fig.~1 for other wavelengths). 
Dashed lines indicate jet directions. 
{\bf b}, Emission centroids of H26$\alpha$ (circles) and H30$\alpha$ (squares) for Source A, 
for channels with peak intensity $>20\sigma$ ($1\sigma=1.8~\mathrm{mJy~beam}^{-1}$). 
The error bars show the uncertainties from channel image fitting.
Offsets are relative to the source continuum peak; ellipses are as in panel a.
{\bf c}, Same as panel b, for H26$\alpha$ of Source B, including channels with $>10\sigma$ ($1\sigma=1.1~\mathrm{mJy~beam}^{-1}$).
{\bf d}, Position change of Source B relative to Source A (zoom in). 
Error bars indicate uncertainties in the relative position offsets 
propagated from the continuum-fitting uncertainties of the two sources.
{\bf e}, R.A. offset between Sources A and B as a function of time. 
Linear fit is shown with blue lines (best-fit highlighted in black).
Black error bars with caps indicate the offset uncertainties propagated from the positional uncertainties from image fitting, 
while the grey shaded bars show uncertainties after accounting for intrinsic scatter in the linear fit.
{\bf f}, Same as panel e, but for Dec. offset.
}\label{fig:proper_motion}
\end{figure}

\begin{figure}
\centering
\includegraphics[width=0.9\textwidth]{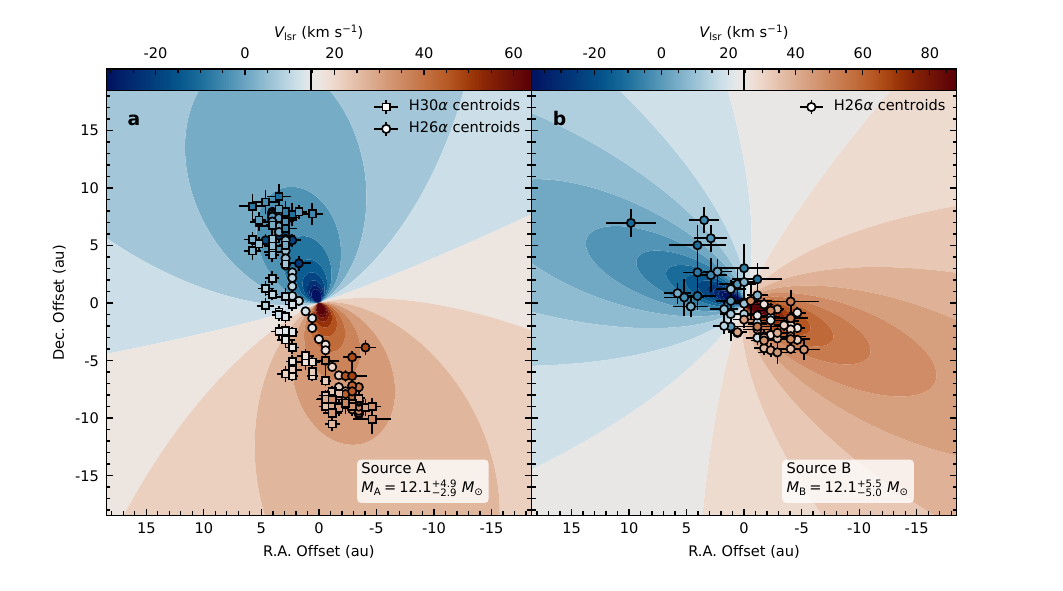}
\caption{\textbf{Keplerian fits to the hydrogen recombination line centroids.}
{\bf a}, Best-fit Keplerian disk model (color map) overlaid with HRL centroids of Source A (H26$\alpha$: circles; H30$\alpha$: squares). The error bars show the positional uncertainties from Gaussian fitting to the channel images.
{\bf b}, Same for Source B (H26$\alpha$ centroids). 
Fitted stellar masses are indicated in the bottom-right corners.
The color scale shows velocity with respect to the local standard of rest ($V_{\rm LSR}$); in each panel, the dashed line indicates the corresponding systemic velocity ($V_{\text{sys,A}}=14.8~\kms$ and $V_{\text{sys,B}}=24.6~\kms$).
The physical scale in au is calculated assuming a distance of 1.68 kpc.
}\label{fig:centroid}
\end{figure}

\begin{figure}
\centering
\includegraphics[width=\textwidth]{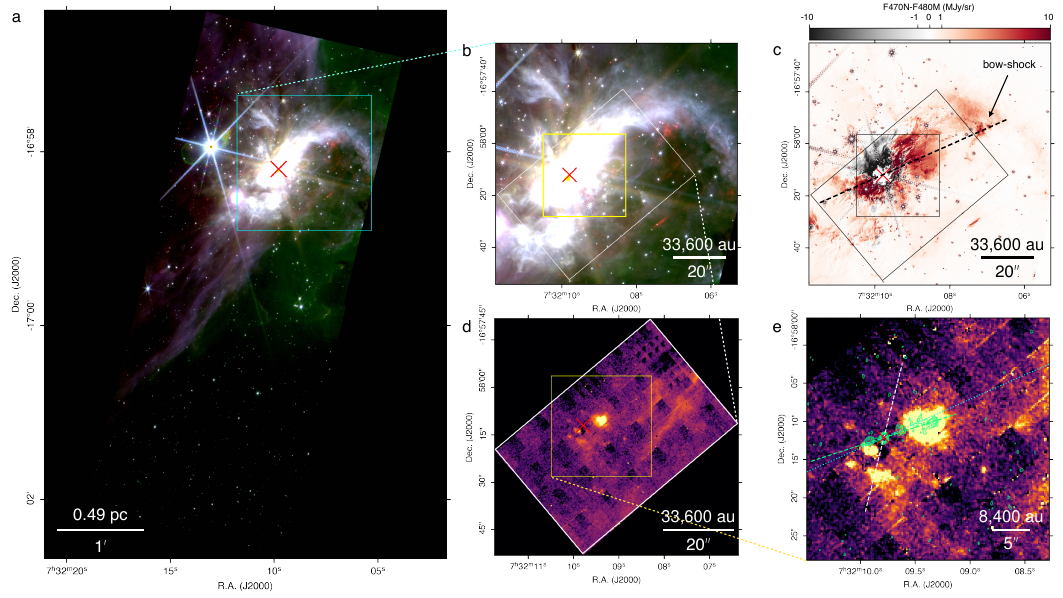}
\caption{
\textbf{JWST/NIRCam and VLT/KMOS images of IRAS 07299$-$1651.}
{\bf a}, JWST false-color image 
(red: H$_2$ 0$-$0 S(9), F470N; green: Br$\alpha$, F405N; blue: 3.6$\mu$m continuum, F360M). 
{\bf b}, Zoom-in with the same color scheme. 
{\bf c}, Continuum-subtracted H$_2$ 0$-$0 S(9) at $4.7~\mu$m (F470N-F480M);
dashed line: source A jet direction, $-66^\circ$.
Positive and negative residuals near the central region in the continuum-subtracted image 
result from imperfect subtraction and are unreliable for identifying jet knots.
{\bf d}, Continuum-subtracted KMOS image of H$_2$ 1-0 S(1) at $2.12~\mu$m.
{\bf e}, Same as panel d with VLA 6~cm contours; dashed lines indicate jet directions 
(green: Source A inner, $-72^\circ$; cyan: Source A large-scale, $-66^\circ$;
white: Source B, $158^\circ$). 
Red cross marks the protobinary. 
The scale bars indicate different physical scales across different panels.
}
\label{fig:IRobs}
\end{figure}

\begin{figure}[ht!]
\centering
\includegraphics[width=0.9\textwidth]{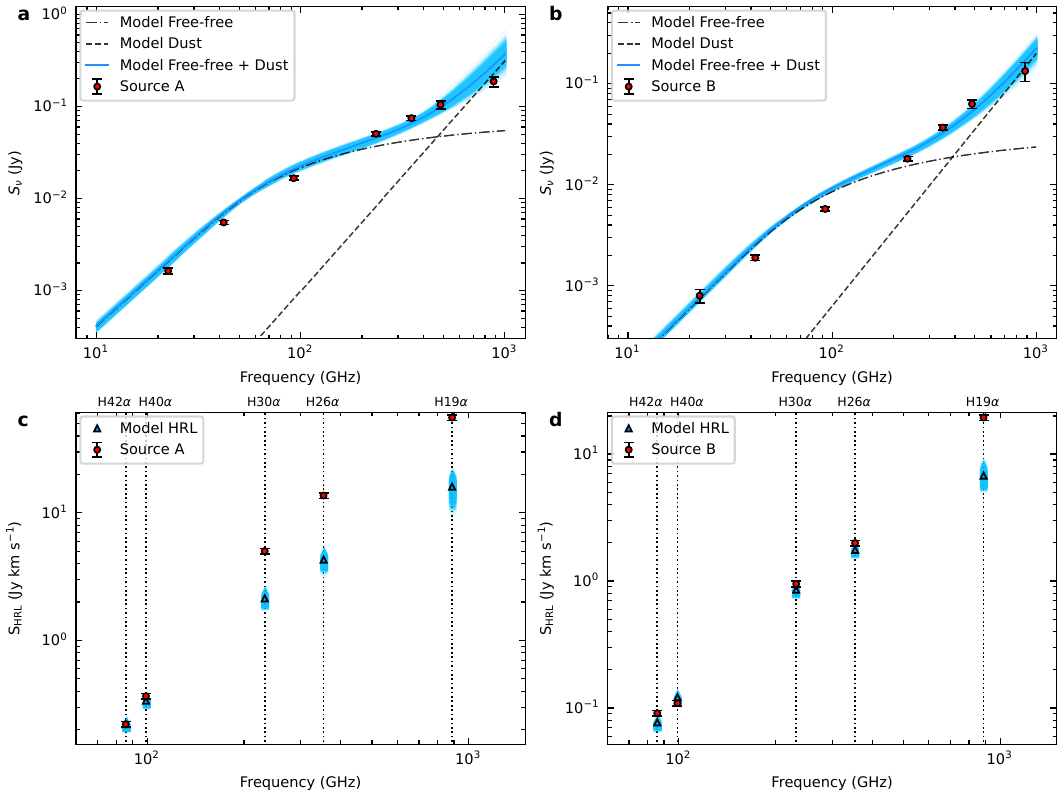}
\caption{\textbf{Joint fitting to the continuum SEDs and HRL fluxes for Source A and B.}
{\bf a,b}, Continuum SEDs of Sources A and B (red points), 
overlaid with the fitted model posterior (light blue lines). 
Error bars combine uncertainties from Gaussian image fitting, 
flux calibration, and additional systematic contributions 
arising from different array configurations and 
the flux differences between the lower and upper sidebands.
The best-fit model, corresponding to the median parameter values, 
is shown as a thick blue line.
The free–free and dust emission contributions are indicated by dot–dashed and dashed lines, 
respectively.
{\bf c,d}, HRL fluxes of Sources A and B (red points), compared with 
model posterior (light blue triangles). 
Error bars combine aperture-summation uncertainties and flux calibration errors of the corresponding bands.
The best-fit model (median parameter values) is highlighted.
Only lower-frequency lines less affected by non-LTE effects are included: 
H42$\alpha$ and H40$\alpha$ for Source A; 
H42$\alpha$, H40$\alpha$, H30$\alpha$, and H26$\alpha$ for Source B.
}\label{fig:sed}
\end{figure}

\begin{figure}[ht!]
\centering
\includegraphics[width=0.9\textwidth]{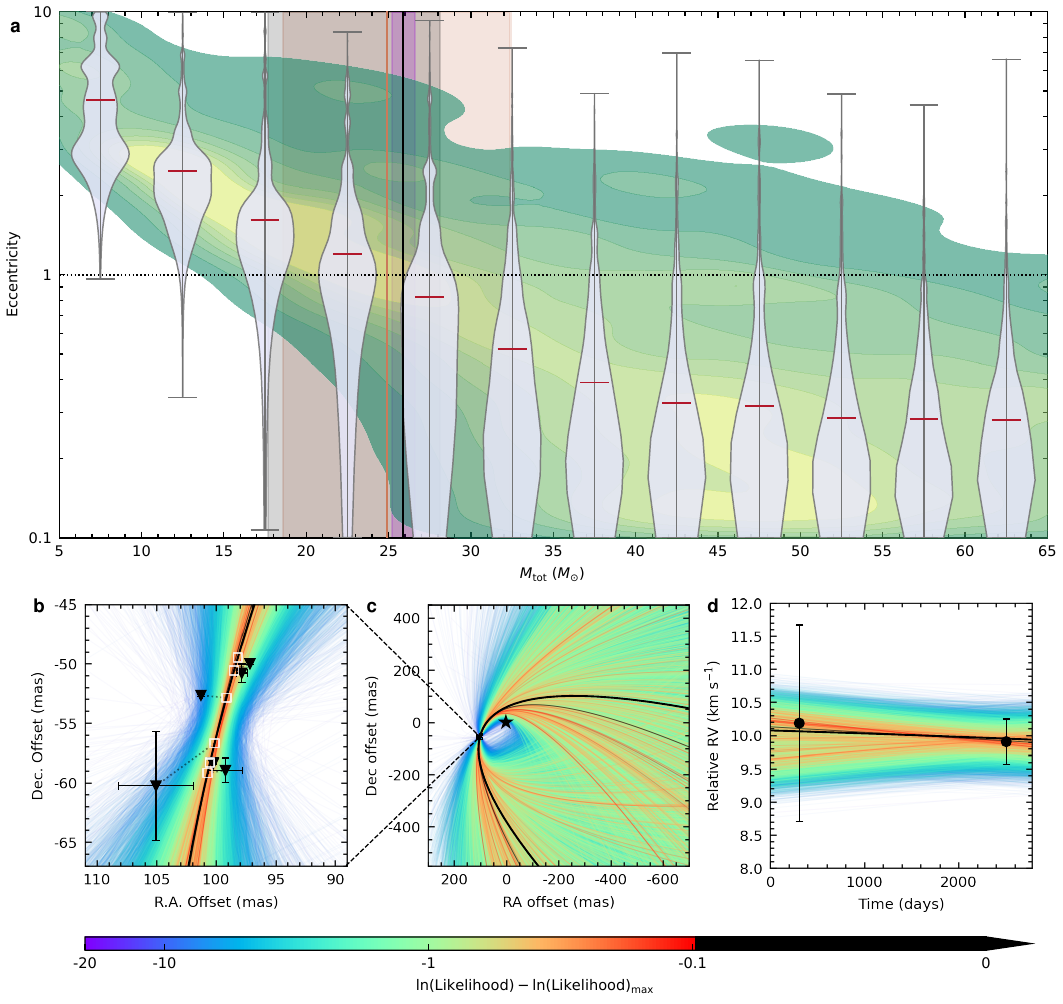}
\caption{\textbf{Orbital constraints on the binary system.}
{\bf a}, Posterior distribution of eccentricity and total system mass 
from the unified orbital fitting (color scale and contours). 
Violin plots show eccentricity distributions in each mass bin, 
with red bars marking medians. 
The black line and purple shaded region indicate the ZAMS mass estimate and uncertainty 
($M_{\rm tot, ZAMS}=25.9\pm0.7M_\odot$) from SED+HRL fitting.
The grey shaded region marks the mass range estimated using protostellar models. 
The brown line and light brown shaded area show the dynamical mass from HRL kinematics 
and its uncertainty.
{\bf b}, Fiducial parabolic fit to the astrometry using constrained mass range (truncated Gaussian prior with $25.9\pm4.0~\msun$ and hard bounds of $17.7$–$28.1~\msun$).
Inverted triangles show the measured positional offsets of Source B relative to Source A.
The error bars of data points indicate uncertainties from continuum image fitting.
Thick black line: best-fit orbit (the maximum likelihood); open squares: model predictions at seven epochs; color scale: difference in $\ln(\mathrm{likelihood})$ from the best-fit orbit.
{\bf c}, Same as panel b, larger view; black star: Source A.
{\bf d}, Same fit applied to the relative radial velocities, shown as filled circles.
The error bars of data points are from the Gaussian fitting to the HRL spectra.
}\label{fig:orbit_para}
\end{figure}

\begin{figure}[ht!]
\centering
\includegraphics[width=0.95\textwidth]{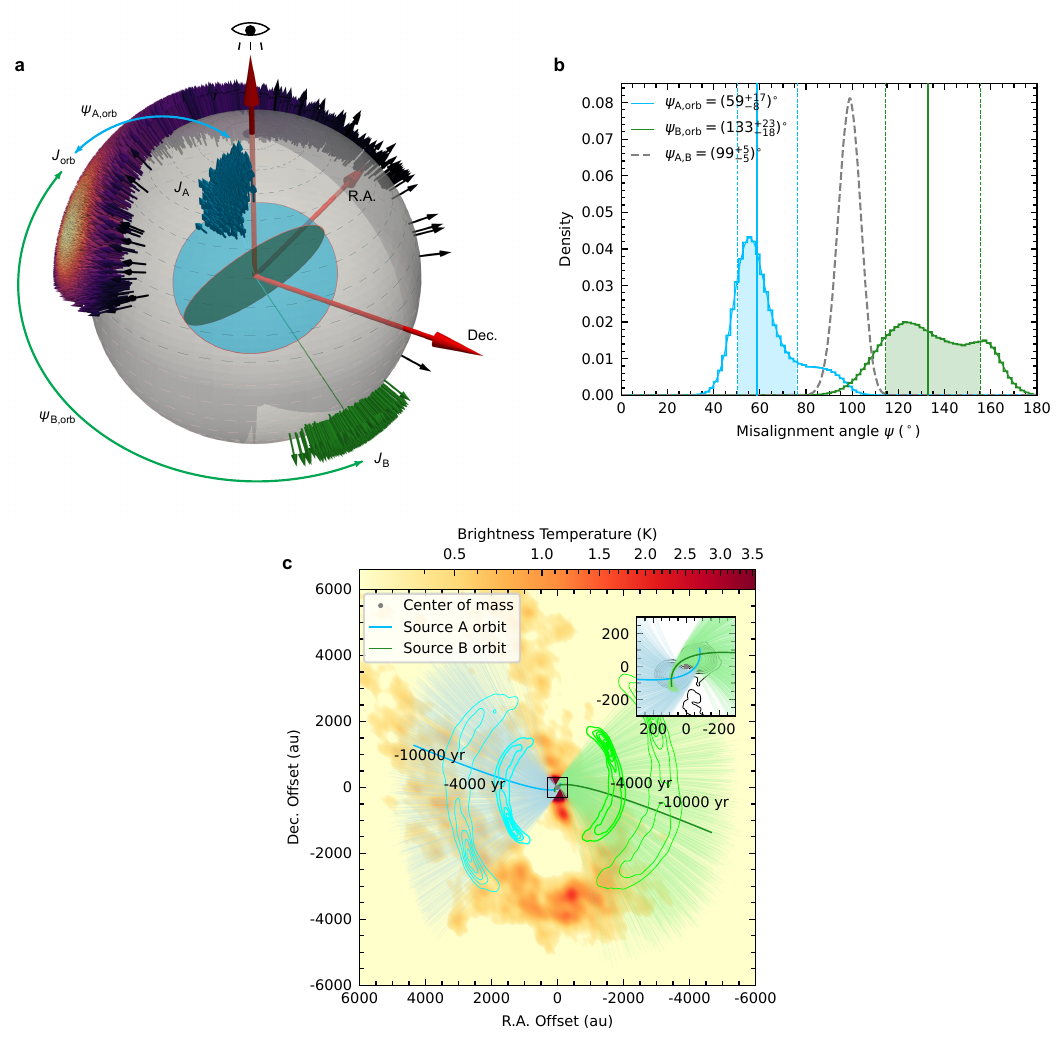}
\caption{\textbf{Disk–orbit misalignment and reconstructed orbital trajectories of the protobinary.}
{\bf a}, Alignment between the circumstellar disks and the orbital plane. 
Angular momentum vectors of the disks ($\mathbf{J}_\text{A}$, $\mathbf{J}_\text{B}$) 
are shown as blue and green arrows. 
Orbital angular momentum vectors $\mathbf{J}_\text{orb}$ from the unified orbital fitting with Gaussian mass prior (Fitting 2) are color-coded by likelihood (brighter: higher likelihood).
{\bf b}, Distribution of inclination angles between the disk angular momenta and the protobinary orbit. 
The shaded regions indicate the central $68\%$ of the probability distribution, 
bounded by vertical dashed lines, while the solid lines mark the median.
The relative angle between $\mathbf{J}_\text{A}$ and $\mathbf{J}_\text{B}$ is shown with a grey dashed line at $\psi_\mathrm{A,B}=99^\circ\pm5^\circ$.
{\bf c}, Past source locations predicted by the fiducial parabolic orbital fitting (Fitting 4), overlaid on the ALMA 1.3 mm~continuum
\cite{zhang_dynamics_2019}, assuming a distance $1.68~\kpc$.
Two large-scale streams are visible: one north of the central sources 
and one curving east from the south. 
Blue and green curves indicate the past trajectories of Sources A and B; 
thick curves mark the best-fit (maximum-likelihood) paths.
Blue/green contours show predicted positions 4000 yr and 10000 yr ago. 
Inset: zoom-in on the central sources.
}\label{fig:misalignment}
\end{figure}

\clearpage

\phantomsection
\label{sec:methods}
\noindent{\large \textbf{Methods}}

\noindent\textbf{ALMA observations and data reduction}

The ALMA observations were obtained at 3~mm on July 6, 2019 
(C10 configuration; baselines $0.19-16.2$ km; project ID 2018.1.01561.S), 
1.3~mm on September 23, 2017 (C9; $0.041-12.1$ km; 2016.1.00125.S), 
0.9~mm on October 10, 14 and 26, 2023 (C8; $0.090-8.5$ km; 2023.1.01721.S), 
0.6~mm on November 3, 2023 (C7; $0.085-8.3$ km; 2023.1.00394.S), 
and 0.3~mm on July 1, 2024 (C6; $0.038-2.5$ km; 2023.1.00394.S). 
The total on-source times were 66, 18, 147, 15 and 27 min 
at 3, 1.3, 0.9, 0.6 and 0.3~mm, respectively.

The data were calibrated using the CASA \cite{casa_team_casa_2022} pipeline 
(v5.4.0 for 3~mm, v4.7.2 for 1.3~mm, v6.5.4 for 0.9 and 0.6~mm, and v6.6.1 for 0.3~mm). 
Each data set was further self-calibrated using continuum data constructed 
from line-free channels, with two phase-only iterations (30~s and 12~s) 
followed by one amplitude iteration with a solution interval of $\sim$1~min. 
The continuum self-calibration solutions were also applied to the HRL data within each band. 
Imaging was performed with the CASA {\it tclean} task 
using Briggs weighting with a robust parameter of 0.5, 
except at 3 mm where a value of $-0.5$ was used to emphasize longer baselines to
achieve comparable angular resolution. 
Synthesized beam sizes and continuum rms noise levels are provided in 
Extended Data Fig.~1.

\vspace{1em}
\noindent\textbf{VLA observations and data reduction}


The VLA 7~mm observations were obtained on 16$-$19 November 2016 in A configuration 
(baselines $0.79-36.6$~km; project 16B-154), with a total on-source time of 34 min. 
The data consist of two 2-GHz wide basebands (3-bit samplers)
centered at 41 and 43~GHz.
The data were calibrated with the CASA pipeline (v4.5.3), 
followed by one phase-only and one amplitude self-calibration iteration. 
Imaging was performed with CASA {\it tclean} using Briggs weighting 
with a robust parameter of $-0.5$ to match the angular resolution of the 1.3~mm continuum.

The VLA 1.3~cm observations were obtained in A configuration on July 4, 2022 
(baselines $0.79-36.6$~km; project 21B-139; 12 min on source) 
and in B configuration on September 19, 2021 
(baselines $0.24-11.1$~km; 57 min on source). 
Both data sets used two 4-GHz basebands (3-bit samplers) centered at 20.45 and 24.45~GHz 
and were calibrated with the CASA pipeline (v6.2.1), 
followed by one phase-only and one amplitude self-calibration iteration. 
Imaging was performed with {\it tclean} using Briggs weighting ($robust=-0.5$). 
The synthesized beam sizes and rms noise levels of the A-configuration continuum are given in Extended Data Fig.~1.
The B-configuration 1.3 cm data are shown in Extended Data Fig.~5.
We also include previously published VLA 6 cm A-configuration data\cite{rosero_soma_2019} to trace the jet.

\vspace{1em}
\noindent\textbf{JWST/NIRCam observations and data reduction}

The JWST observations were obtained on March 11, 2024 (Programm ID:~3907; PI:~Y. Zhang).
The Near-Infrared Camera (NIRCam)\cite{rieke05} was employed using the FULLBOX primary dither pattern with six TIGHT dithers, providing a total of six individual pointings. This configuration resulted in a final field of view (FoV) of $\sim2^{\prime}\times 6^{\prime}$.
Paired short- and long-wavelength filters (F162M/F405N, F212N/F470N, F182M/F480M and F115W/F360M) were observed using the \textit{SHALLOW2} readout pattern. 
Most filters used six groups per integration and one integration per exposure, yielding a total exposure time of 1,739~s; the F115W/F360M pair used three groups per integration for a total exposure time of 773~s. The position angle relative to the V3 axis (\texttt{PA\_V3}) was $76.13^\circ$.

Data reduction was carried out with the JWST calibration pipeline (v1.12.5)\citep{jwst_calib_1_12_5}
using the Calibration Reference Data System (CRDS) context \texttt{jwst\_1217.pmap}. 
Standard Stage~1$-$3 reduction steps were executed with minor modifications. 
In Stage~1, \texttt{suppress\_one\_group = False} was set in the \texttt{ramp\_fit} step to recover saturated pixels between the first and second reads. 
After Stage~2, the 1/$f$ noise was mitigated using the \texttt{image1overf.py} script.
In Stage 3, a common output frame was enforced across all filters, with final image dimensions of $8,964\times12,204$~pixels and a uniform pixel scale of $0.031^{\prime\prime}$, 
corresponding to the native sampling of the short-wavelength channel. 
All images were finally rotated to standard astronomical orientation, with north up and east to the left.

\vspace{1em}
\noindent\textbf{VLT/KMOS observations and data reduction}

The VLT K-band Multi Object Spectrograph (KMOS)
\citep{kmos1,kmos2} observations were obtained on December 17, 2021 and January 21, 2022 (Program ID:~108.223D.001, PI:~R.~Fedriani). 
KMOS consists of 24 integral field units (IFUs), each providing a FoV of $2.8^{\prime\prime}\times2.8^{\prime\prime}$,
and in mosaic mode, 16 pointings covering $\sim 43.4^{\prime\prime}\times64.8^{\prime\prime}$. 
Each KMOS pointing was observed for 180~s, resulting in a total integration time of 2880~s. 
The spatial sampling is $0.2^{\prime\prime}\times0.2^{\prime\prime}$, with a seeing-limited angular resolution of $0.7^{\prime\prime}-1.4^{\prime\prime}$. 
The spectral resolving power in the $K$ band is $R\sim4200$ (i.e. a spectral resolution of $\sim$0.52~nm, or $\sim70~\kms$).
The data were processed using the ESOReflex KMOS pipeline \citep{esoreflex}. 
Astrometric calibration was performed using 2MASS field stars,
and the \texttt{reproject} package, together with the function 
\texttt{find\_optimal\_celestial\_wcs}, 
was used to orient the images with north up and east to the left.

\vspace{1em}
\noindent\textbf{Continuum deconvolution and position determination}

We performed two-dimensional, two-component Gaussian fitting to the continuum images 
using the CASA task {\it imfit} (Supplementary Fig.~1). 
The fits yield the source positions, peak intensities, and integrated flux densities 
at each wavelength. 
As the {\it imfit} procedure assumes that the observed emission is the convolution of 
the synthesized beam with an intrinsic Gaussian component, 
the fits also provide deconvolved source morphologies, 
including major and minor axis full widths at half-maximum (FWHM) and position angles. 
These parameters are summarized in Supplementary Table~1. 
The deconvolved Gaussian components are shown as ellipses 
in Fig.~1a and Extended Data Fig.~4. 
The fitting residuals are generally below the $3\sigma$ level (Supplementary Fig.~1).

To further assess the robustness of the image-domain results, 
we directly fit the visibility data with two Gaussian components using the Python package {\it UVMultiFit} \cite{uvmultifit2014}. 
Owing to the low signal-to-noise ratio of the 13 mm data and the marginally resolved nature of the sources, 
we adopt uniform weighting and fix the Gaussian profiles to be circular to obtain stable fits. 
All fits are satisfactory (Supplementary Fig.~2) 
and yield results consistent with those from {\it imfit} (Supplementary Fig.~3).
The reliability of continuum deconvolution and position determination is discussed in Discussion 1 and 2 in Supplementary Information.

\vspace{1em}
\noindent\textbf{Determination of HRL central velocities and radial velocity offsets}

We fitted single-component Gaussian profiles to all HRL spectra 
extracted at the peak positions in the corresponding continuum images of both sources, 
except for the weak H42$\alpha$ and H40$\alpha$ lines in Source B. 
Although the H26$\alpha$ line in Source A exhibits a double-peaked profile, 
commonly associated with disk rotation and potentially enhanced by maser amplification \cite{zhang_angular_2017} 
(see Discussion 3 in Supplementary Information), 
we still adopt a single-component Gaussian fit to this line,
solely to determine the systemic velocity, 
deferring detailed kinematic analysis to the ``Fitting HRL centroids'' section.
The resulting HRL central velocities and line widths are shown in 
Extended Data Fig.~3 and listed in Supplementary Table~2. 

In both sources, the central velocities of the H26$\alpha$ and H30$\alpha$ transitions are consistent. 
The H42$\alpha$ and H40$\alpha$ lines in Source A show slightly lower central velocities but 
remain consistent with the H30$\alpha$ velocity within $2\sigma$. 
By contrast, the H19$\alpha$ transitions yield systematically different central velocities 
from those measured in the lower-$n$ lines. 
Because the H19$\alpha$ transition may be affected by dust opacity and non-LTE effects (see Discussion 3 in Supplementary Information), 
we do not use it to determine the source radial velocities. 
We therefore adopt the H26$\alpha$ and H30$\alpha$ central velocities as the line-of-sight velocities 
of the two sources and use their differences as the radial velocity offsets at the corresponding epochs.


\vspace{1em}
\noindent\textbf{Fitting HRL centroids}

For the H30$\alpha$ line (Source A only) and the H26$\alpha$ line (Sources A and B), 
we used CASA {\it imfit} task to perform two-dimensional Gaussian fits to individual channels to 
determine the emission centroids, using channels with peak S/N$>20$ for Source A 
and $>10$ for Source B (Fig.~1b,c).

We model the centroid kinematics using a simple Keplerian disk rotation model
\cite{cesaroni_dissecting_2025},
parameterized by the central mass $M$,
systemic velocity $V_{\text{sys}}$,
disk inclination $i_d$ (relative to the plane of the sky), disk position angle $\mathrm{PA}_d$,
and distance to this system $d$.
For each velocity channel $v_i$ and a given set of model parameters, 
we compute the predicted line-of-sight velocity $v_{\mathrm{model},i}$
at the observed centroid position assuming Keplerian rotation,
and minimize the residual
\begin{equation}
  \sum_i \frac{v_i-v_{\text{model},i}}{\Delta v_{\text{model},i}}, 
\end{equation}
where $\Delta v_{\text{model},i}$ is the uncertainty 
in the model velocity arising from the centroid position error.
The summation is over all channels with reliable centroid measurements. 
The fitting is performed using the Python package \texttt{lmfit} \cite{lmfit_software}. 
In the fitting, the systemic velocities are fixed to the central velocities of the corresponding HRLs. 
Because the primary goal of the centroid analysis is 
to test whether the emission is consistent with ionized Keplerian circumstellar disks, 
we further fix the disk inclinations and position angles 
to the values derived from the continuum data
(allowing variations within uncertainties). 
With these constraints, the only free parameter is the central mass.

For Source A, centroid measurements from both the H30$\alpha$ and H26$\alpha$ lines are used, 
whereas for Source B only the H26$\alpha$ line is included owing to its higher signal-to-noise ratio. 
To propagate the uncertainties in the fixed geometrical and physical parameters into the final dynamical mass estimates, 
we adopt a Monte Carlo approach. 
We generate 1000 realizations of the fixed parameters, drawing the inclination $i_\text{d}$ and the system distance $d$ 
independently from Gaussian distributions defined by their measured uncertainties
(e.g., $\sigma_{i_{\text{d}}}=5^{\circ}$ and $\sigma_{d}=0.1~\kpc$). 
For each realization, we re-optimize the model to obtain the best-fit central mass and its associated uncertainty. 
The resulting 1000 sub-posteriors are then combined to construct the final posterior distribution, 
from which the central mass and its uncertainty are derived.
The fitting results are shown in Fig.~2. 

\vspace{1em}
\noindent\textbf{Fitting multi-wavelength continuum and HRL fluxes}

The continuum spectral energy distributions (SEDs) of the two sources were constructed 
from flux densities measured via two-component Gaussian fitting to the continuum images (see ``Continuum deconvolution and position determination''). 
Uncertainties were calculated by combining the Gaussian-fitting uncertainties with flux calibration uncertainties, 
adopting 5\% for the VLA data and ALMA Bands 3, 6, and 7, and 10\% for ALMA Bands 8 and 10. 
Additional uncertainty terms were included to account for the flux differences between ALMA images 
with and without compact-configuration data in Bands 7 and 10, 
and for half of the lower–upper baseband flux difference in the VLA K band. 
Hydrogen recombination line (HRL) fluxes were measured by integrating HRL emission within 
elliptical apertures (Extended Data Fig.~3).
Their uncertainties were computed by combining the aperture integration uncertainties
with the same flux calibration uncertainties adopted for the continuum measurements.

We model the continuum SEDs and HRL fluxes of both sources 
using a simple ionized disk model. 
The model assumes a power-law radial distribution of emission measure (EM),
$\text{EM}(r)=\text{EM}_0(r/r_0)^{-p}$. 
Here $r$ is the disk radius, $\text{EM}_0$ is the emission-measure normalization at the reference radius $r_0=10~\au$, and $p$ is the power-law index.
The free-free optical depth is given by \cite{draine_physics_2011}
\begin{equation}
\tau_{\nu,\text{ff}} = 3.366\times10^{-7} \left(\frac{T}{10^4~\text{K}}\right)^{-1.323}\left(\frac{\nu}{10^9~\text{Hz}}\right)^{-2.118} 
\left(\frac{\elecm}{\pccm}\right),
\end{equation}
where $T$ is the ionized gas temperature and $\nu$ is the observing frequency.
Assuming a geometrically thin disk with finite thickness and an inclination $i$
between the disk axis and the line of sight, the free–free flux density is
\begin{equation}
    S_{\nu,\text{ff}} = \frac{2\pi  B_{\nu}(T)}{d^2}\cos i \int_{r_{\text{in}}}^{r_{\text{out}}} \left(1 - e^{-\frac{ \tau_{\nu,\text{ff}}(r)}{\cos i}}\right) r\mathrm{d}r,
\end{equation}
where $B_\nu(T)$ is the Planck function, $d$ is the source distance and $r_\mathrm{in}$ and $r_\mathrm{out}$ are the inner and outer radii of the ionized disk.

An electron density gradient, encapsulated by the EM power law, 
is required because a uniform-density ionized region would produce a sharp transition 
between optically thick and optically thin regimes in the SED, which is not observed. 
An outer truncation radius $r_\mathrm{out}$ is also necessary; 
without it, reproducing the low-frequency flux densities would require 
unrealistically low gas temperatures. 

At the highest frequencies, dust emission begins to dominate the continuum, 
we therefore include a dust contribution parameterized as a single power law,
\begin{equation}
S_{\nu,\text{dust}}=S_{0.3\text{mm},\text{dust}}(\frac{\nu}{\nu_{0.3\text{mm}}})^{-\alpha_{\text{d}}}. 
\end{equation}
Here $S_{0.3\text{mm},\text{dust}}$ is the dust flux density at the reference frequency $\nu_{0.3\text{mm}}$ corresponding to 0.3~mm, and $\alpha_\mathrm{d}$ is the dust spectral index.

For hydrogen recombination line emission  H$n\alpha$  (the $n+1 \to n$ transition), 
the line optical depth is given by \cite{gordon_radio_2002}
\begin{equation}
\tau_{\nu,\mathrm{H}n\alpha} \approx 2.042\times 10^{6} \phi_{\nu}  T^{-5/2} \left(1-\frac{9}{4n^2}\right) \exp\left(\frac{E_n}{kT}\right) \left(\frac{\elecm}{\pccm}\right) \left(\frac{N(\text{H}^{+})}{N(\text{He}^{+})+N(\text{H}^{+})}\right),
\end{equation}
where $\phi_\nu$ is the normalized line profile, $E_n$ is the ionization energy of hydrogen from level $n$, $k$ is the Boltzmann constant, and $N(\text{H}^{+})$ and $N(\text{He}^{+})$ are the number densities of H$^+$ and He$^+$, respectively.
The final factor accounts for the fact that the relevant ions arise from H$^+$, whereas free electrons originate from both H$^+$ and He$^+$.
We adopt $N(\text{He}^{+})/N(\text{H}^{+})=0.08$ \cite{shaver_galactic_1983}.
The continuum-subtracted line intensity at radius $r$ is 
\begin{equation}
I_{\nu,\mathrm{H}n\alpha}(r)=B_{\nu}(T)
\left(1-e^{-\frac{\tau_{\nu,\mathrm{H}n\alpha}(r)}{\cos i} }\right)e^{-\frac{\tau_{\nu,\text{ff}}(r)}{\cos i}},
\end{equation} 
and the total line flux integrated over velocity and emitting area, is 
\begin{equation}
S_{\mathrm{H}n\alpha} = \frac{2\pi\cos i}{d^2} \int_{r_{\text{in}}}^{r_{\text{out}}}r \mathrm{d}r\int B_{\nu}(T)\left(1-e^{-\frac{\tau_{\nu,\mathrm{H}n\alpha}(r)}{\cos i} }\right)
e^{-\frac{\tau_{\nu,\text{ff}}(r)}{\cos i}}\mathrm{d} \nu \left(\frac{c}{\nu_0}\right),
\end{equation}
where $\nu_0$ is the rest frequency of the $\mathrm{H}n\alpha$ transition. 
Here $c$ is the speed of light, and the factor $c/\nu_0$ converts the frequency-integrated flux density to units of 
$\Jykms$, such that $S_{\mathrm{H}n\alpha}$ retains an explicit dependence on $\nu_0$.

To reduce the number of free parameters, the ionized gas temperature is fixed at a typical value of 8,000~K \cite{keto_early_2008}.
The inner radius of the disk is set to $r_{\text{in}}=0.05~\text{au}$, 
approximately the stellar radius. 
The dust spectral index is fixed at $\alpha_\mathrm{d}=2.5$, 
an intermediate value between optically thin and optically thick dust emission.

The model therefore contains six free parameters: 
the emission measure normalization 
$\text{EM}_0$ at the reference radius $r_0=10~\au$, 
the power-law index of the EM distribution $p$, 
the outer truncation radius of the ionized disk $r_{\text{out}}$, 
the disk inclination angle $i$, 
the dust continuum normalization at 0.3 mm, $S_{0.3\text{mm},\text{dust}}$,
and the source distance $d$. 
The distance $d$ is treated as a free parameter to account for 
the impact of its uncertainty on the fitted results. 
A Gaussian prior of $1.68\pm 0.1~\kpc$ is adopted for $d$\cite{distance1680}.
We simultaneously fit the total continuum flux density, 
$S_{\nu,\text{ff}} + S_{\nu,\text{dust}}$, 
and the velocity-integrated hydrogen recombination line fluxes, $S_{\text{HRL}}$, 
using a Markov Chain Monte Carlo (MCMC) approach \cite{foreman-mackey_emcee_2013}.
For posterior distributions that have converged -- 
as assessed using the Gelman–Rubin diagnostic \cite{GRtest1992,arviz_2019} -- 
the full sample is used to derive the ionizing photon rates and stellar masses, 
allowing uncertainties to propagate naturally.
The corner plots of fitting posterior samples for Sources A and B 
are shown in Supplementary Figs.~4 and 5, respectively. 
The median values and corresponding $68\%$ credible intervals are listed in Extended Data Table~1.

We note that because the high-frequency HRLs show evidence for non-LTE effects (see Discussion 3 in Supplementary Information), 
we fit only the lower-frequency HRL fluxes, where non-LTE effects are unlikely to be substantial. 
For Source A, H42$\alpha$ and H40$\alpha$ are included, 
while for Source B, H42$\alpha$, H40$\alpha$, H30$\alpha$ and H26$\alpha$ are included.

\vspace{1em}
\noindent\textbf{Fitting orbital motions}

Assuming that the relative proper motions and radial velocity differences 
of the two sources arise from their orbital motion, 
we fit these quantities simultaneously using a two-body dynamical model. 
The governing equation is \cite{murray_keplerian_2010}:
\begin{equation}
\ddot{\mathbf{r}} + \frac{G M_{\mathrm{tot}}}{r^3} \mathbf{r} = 0,
\end{equation}
where $\mathbf{r}$ is the relative position vector of Source B with respect to Source A, $G$ is the gravitational constant, and $M_{\mathrm{tot}}=M_{\mathrm{A}}+M_{\mathrm{B}}$ is the total system mass.
Depending on the eccentricity, the orbit may be elliptical ($e<1$), parabolic ($e=1$) or hyperbolic ($e>1$), corresponding to bound, critical and unbound solutions.
As $e\to 1$, both elliptical and hyperbolic orbits converge to the parabolic case, 
for which the semi-major axis $a$ diverges.
To avoid this singularity, we adopt the periapsis distance 
$q=a\left(1-e\right)$
as the fitting parameter instead of $a$.
All three orbital families are therefore described within a common parameter space 
consisting of the periapsis distance $q$, eccentricity $e$, 
longitude of the ascending node $\Omega$ (measured east of north), 
argument of periapsis $\omega$,
inclination $i$ 
of the orbital plane relative to the plane of the sky, 
time of periapsis passage $t_{\mathrm{p}}$ (MJD) and total system mass $M_{\text{tot}}$. 
To account for additional, uncharacterized astrometric uncertainties, 
we include an intrinsic scatter term, $\sigma_{\text{int}}$ in the fitting\cite{Hogg10}. 
The distance, $d=1.68\pm0.1~\kpc$ \cite{distance1680},
is also treated as a free parameter to incorporate its uncertainty into the analysis.

For this unified orbital model with nine free parameters, the likelihood for a given parameter vector $\boldsymbol{\Theta}$ is given by:
\begin{eqnarray}
\mathcal{L}(\boldsymbol{\Theta}) & = & \exp \left\{
\frac{N_{\alpha}+N_{\delta}+N_{v}}{3} 
\left(
\frac{\ln\mathcal{L}_\alpha}{N_\alpha}+\frac{\ln\mathcal{L}_\delta}{N_\delta}+\frac{\ln\mathcal{L}_v}{N_v}
\right)
\right\},\nonumber\\
\ln\mathcal{L}_\alpha & = & -\frac{1}{2}\sum_{i} \left( \frac{[\Delta \alpha_i - \Delta \alpha(\boldsymbol{\Theta} ,t^a_i)]^2}{\sigma^2_{\Delta \alpha,i}+\sigma^2_{\text{int}}} + \ln\left(2\pi\left[\sigma^2_{\Delta \alpha,i}+\sigma^2_{\text{int}}\right]\right) \right) \notag, \\
\ln\mathcal{L}_\delta & = & -\frac{1}{2}\sum_{i} \left( \frac{[\Delta \delta_i - \Delta \delta(\boldsymbol{\Theta} ,t^a_i)]^2}{\sigma^2_{\Delta \delta,i}+\sigma^2_{\text{int}}} + \ln\left(2\pi\left[\sigma^2_{\Delta \delta,i}+\sigma^2_{\text{int}}\right]\right) \right) \notag, \\
\ln\mathcal{L}_v & = & -\frac{1}{2} \sum_{j} \left( \frac{[\Delta v_{\mathrm{los},j} - \Delta v_{\mathrm{los}}(\boldsymbol{\Theta} ,t^v_j)]^2}{\sigma^2_{\Delta v_{\mathrm{los},j}}} + \ln\left(2\pi\sigma^2_{\Delta v_{\mathrm{los},j}}\right) \right).
\label{eq:likelihood}
\end{eqnarray}
Here $t^{a}_i$ and $t^{v}_j$ denote the epochs of the astrometric and radial velocity measurements, respectively.
$\Delta \alpha(\boldsymbol{\Theta},t)$, $\Delta \delta(\boldsymbol{\Theta},t)$, and $\Delta \vlos(\boldsymbol{\Theta},t)$ are
the model predictions for the position offsets and radial velocity offsets 
given parameters $\boldsymbol{\Theta}$ and epoch $t$.
$\Delta \alpha_i$, $\Delta \delta_i$, and $\Delta v_{\mathrm{los},j}$ are the observed values
with uncertainties
$\sigma_{\Delta \alpha,i}$, $\sigma_{\Delta \delta,i}$, and $\sigma_{\Delta v_{\mathrm{los},j}}$. 
$N_\alpha$, $N_\delta$, and $N_v$ are the numbers of R.A.-offset, Dec.-offset, and radial-velocity data points, respectively.
The three components are weighted inversely by their respective numbers of data points, 
ensuring that astrometry and radial velocity contribute roughly equally to the likelihood.
The posterior distribution is $p(\boldsymbol{\Theta} )\propto\mathcal{L}(\boldsymbol{\Theta} )\pi(\boldsymbol{\Theta})$, 
where $\pi(\boldsymbol{\Theta})$ is the prior distribution.

To evaluate the goodness-of-fit, we fixed $\sigma_{\mathrm{int}}$ to the best-fit value (median) and calculated the weighted reduced chi-square by definition ($N_\mathrm{tot}=N_\alpha+N_\delta+N_v$ is the total number of data points, $N_{\boldsymbol{\Theta}}$ is the number of free model parameters, and $N_\mathrm{dof}=N_\mathrm{tot}-N_{\boldsymbol{\Theta}}$):

\begin{equation}
\begin{aligned}
\chi^2_{\text{reduced}} &= \frac{N_{\mathrm{tot}}}{3 N_{\mathrm{dof}}} \Bigg[ 
  \frac{1}{N_\alpha} \sum_{i=1}^{N_\alpha} \frac{[\Delta \alpha_i - \Delta \alpha(\boldsymbol{\Theta}, t^a_i)]^2}{\sigma^2_{\Delta \alpha,i} + \sigma^2_{\mathrm{int}}} \\
  &\quad + \frac{1}{N_\delta} \sum_{i=1}^{N_\delta} \frac{[\Delta \delta_i - \Delta \delta(\boldsymbol{\Theta}, t^a_i)]^2}{\sigma^2_{\Delta \delta,i} + \sigma^2_{\mathrm{int}}} \\
  &\quad + \frac{1}{N_v} \sum_{j=1}^{N_v} \frac{[\Delta v_{\mathrm{los},j} - \Delta v_{\mathrm{los}}(\boldsymbol{\Theta}, t^v_j)]^2}{\sigma^2_{\Delta v_{\mathrm{los},j}}} 
\Bigg].
\end{aligned}
\label{eq:reduced_chi2_full}
\end{equation}



We first perform an exploratory fit (Fitting 1) to sample a broad region of parameter space. 
In this fit, the eccentricity $e$ is assigned a log-uniform prior over $0.1 - 10$,
and the total mass $M_{\text{tot}}$ a uniform prior over $0-100~M_{\odot}$.
The distance $d$ is given a Gaussian prior with $\mu_d=1.68~\kpc$ and $\sigma_d=0.1~\kpc$.
The remaining parameters are assigned uniform priors:
periapsis distance $q$ over $0-1000~\au$, 
inclination $i$ over $0-\pi$, 
argument of periapsis $\omega$ over $0-2\pi$, 
longitude of ascending node $\Omega$ over $0-2\pi$, 
time of periapsis $t_{\mathrm{p}}$ from 0 to twice the band 10 observation date, 
and intrinsic scatter $\sigma_{\text{int}}$ over $0-3$ mas.
In the fitting, orbits are classified as elliptical for $e<1-\epsilon$, hyperbolic for $e>1+\epsilon$, 
and parabolic for $1-\epsilon\le e\le1+\epsilon$,
with $\epsilon=10^{-4}$.
Elliptical orbits are computed using the \texttt{orbitize.kepler.calc$\_$orbit} function
in the \texttt{orbitize!} package 
\cite{blunt_orbitize_2020,blunt_orbitize_2023,blunt_first_2023},
while parabolic and hyperbolic orbits are computed with the \texttt{pyorb} package . 
We sample the posterior using a parallel-tempered Markov Chain Monte Carlo (MCMC) 
algorithm \cite{vousden_dynamic_2016} with 100 walkers per temperature and 30 temperatures, 
running 50,000 burn-in steps followed by 50,000 production steps. 
Convergence was confirmed via the Gelman–Rubin diagnostic, 
and the resulting corner plot is shown in Supplementary Fig.~6. 

We perform a second orbital fit (Fitting 2) with the total mass constrained by independent estimates. 
Specifically, we adopt a truncated Gaussian prior for $M_\text{tot}$, with mean $25.9~\msun$, standard deviation $4.0~\msun$, and hard bounds of $17.7$–$28.1~\msun$.
A uniform prior over the range $0-2$ is adopted on the eccentricity.
The resulting corner plot is shown in Supplementary Fig.~7.
We further perform elliptical orbital fitting using the 1.3 mm data alone 
(i.e., radial velocity difference and single-epoch astrometry), 
similar to the previous study of this source\cite{zhang_dynamics_2019} 
(Fitting 3; Discussion 4 in Supplementary Information; Supplementary Fig.~8).
As the fiducial model (Fitting 4), we adopt a parabolic orbit ($e=1$) with the total mass constrained to $M_\text{tot}=25.9\pm4.0~\msun$ and hard bounds of $17.7$–$28.1~\msun$ (Supplementary Fig.~10).
The resulting orbital solutions are shown in Fig.~5b–d, with the best-fit (maximum-likelihood) orbit 
 highlighted in black.
A summary of the four orbital fits with different prior assumptions and datasets 
is provided in Extended Data Table~2.

\vspace{1em}
\noindent\textbf{Alignments between circumstellar disks and orbital plane}

The angular momentum vectors of the two circumstellar disks and the binary orbit, 
$\mathbf{J}_\text{A}$, $\mathbf{J}_\text{B}$ and $\mathbf{J}_\text{orb}$, 
are parameterized by their inclination and position angles: 
$(i_\mathrm{A}, \phi_\mathrm{A})$, $(i_\mathrm{B}, \phi_\mathrm{B})$, and $(i_\mathrm{orb}, \phi_\mathrm{orb})$, respectively.
Here, the inclination is defined such that $i>0^{\circ}$ when the angular momentum vector points toward the observer, 
and $\phi$ denotes the position angle of its projection onto the plane of the sky, measured from north to east.
Notably, $\phi$ is simply $90^{\circ}$ less than the measured position angle (P.A.) of the disk major axis 
(for circumstellar disks) the longitude of the ascending node (for the orbit). 
The relative misalignment angle, $\psi_{\alpha\beta}$, between any two angular momentum vectors is given by
\begin{equation}
    \cos \psi_{\alpha\beta} = \sin i_{\alpha} \sin i_{\beta} \cos(\phi_{\alpha} - \phi_{\beta}) + \cos i_{\alpha} \cos i_{\beta},
\end{equation}
where $\alpha$ and $\beta$ denote either of the two circumstellar disks or the binary orbit.

From the orbital fitting, we obtain posterior samples of the orbital inclination $i_\mathrm{orb}$ and 
longitude of the ascending node $\Omega$, 
which are transformed into $(i_\mathrm{orb}, \phi_\mathrm{orb})_i$. 
We then draw corresponding samples of $(i_\mathrm{A}, \phi_\mathrm{A})_i$ and $(i_\mathrm{B}, \phi_\mathrm{B})_i$ 
from their respective uncertainty distributions. 
These realizations are used to compute the posterior distributions of the relative misalignment angles.
While the magnitudes of the inclination angles of the circumstellar disks are relatively well constrained, 
their signs remain poorly determined. 
We show the misalignment angle distributions for the most likely configuration 
(inferred from infrared jet observations; see ``Jet and disk orientations'') 
in Fig.~6, and for the other three configurations in Extended Data Fig.~7.

\vspace{1em}
\noindent\textbf{Data Availability}

This paper makes use of the following ALMA data: ADS/JAO.ALMA \#2016.1.00125.S, \#2018.1.01561.S, \#2023.1.00394.S, and \#2023.1.01721.S,
the following JVLA data: \#16B-154 and \#21B-139, the following JWST data: \#3907, and the following VLT data: \#108.223D.001.
The raw ALMA data are available at \url{https://almascience.nao.ac.jp/aq}.
The raw JVLA data are available at \url{https://data.nrao.edu}.
The raw JWST data are available at the Mikulski Archive for Space Telescopes at the Space Telescope Science Institute (\url{https://mast.stsci.edu/}).
The raw VLT data are available at the ESO data archive (\url{http://archive.eso.org/}). 
The reduced ALMA and VLA continuum images and hydrogen recombination line image cubes
used in this study have been deposited in Zenodo\cite{wang_reduced_2026} 
and are available at \url{https://doi.org/10.5281/zenodo.20913729}.

\vspace{1em}
\noindent\textbf{Code Availability}

The ALMA and VLA data were reduced using the corresponding pipeline versions of CASA, 
and imaged using CASA v6.6.4 (\url{https://casa.nrao.edu/casa_obtaining.shtml}). 
The \texttt{image1overf.py} script used to mitigate the 1/$f$ noise in the JWST images is publicly available at \url{https://github.com/chriswillott/jwst/blob/master/image1overf.py}.
The model orbits were generated using the open-source Python packages \texttt{orbitize!} v3.1.0\cite{blunt_orbitize_2020,blunt_orbitize_2023} (\url{https://github.com/sblunt/orbitize}) 
and \texttt{pyorb} v0.6.0 (\url{https://github.com/danielk333/pyorb}).
The HRL spectral fitting and centroid kinematics fitting were performed with the package \texttt{lmfit} v1.3.2\cite{lmfit_software} (\url{https://github.com/lmfit/lmfit-py}). 
The visibility data are fitted with the package \texttt{UVMultiFit} v3.1\cite{uvmultifit2014} (\url{https://github.com/marti-vidal-i/UVMultiFit}).
The posterior distributions in SED+HRL fitting and orbital fitting were sampled using 
the package \texttt{emcee} v3.1.6\cite{foreman-mackey_emcee_2013} (\url{https://github.com/dfm/emcee}). 
To evaluate the convergence of posterior samples, we utilized the package \texttt{arviz} v0.17.1\cite{arviz_2019} (\url{https://github.com/arviz-devs/arviz}).

\vspace{1em}
\noindent\textbf{Acknowledgements}

This paper makes use of the following ALMA data: \#2016.1.00125.S, \#2018.1.01561.S, \#2023.1.00394.S, and \#2023.1.01721.S. 
ALMA is a partnership of ESO (representing its member states), NSF (USA) and NINS (Japan), 
together with NRC (Canada), NSTC and ASIAA (Taiwan), and KASI (Republic of Korea), 
in cooperation with the Republic of Chile. 
The Joint ALMA Observatory is operated by ESO, AUI/NRAO and NAOJ. 
The National Radio Astronomy Observatory and Green Bank Observatory are facilities of the U.S. National Science Foundation operated under cooperative agreement by Associated Universities, Inc.
This work is based in part on observations made with the NASA/ESA/CSA James Webb Space Telescope. 
The data were obtained from the Mikulski Archive for Space Telescopes at the Space Telescope Science Institute, 
which is operated by the Association of Universities for Research in Astronomy, Inc., 
under NASA contract NAS 5-03127 for JWST. 
These observations are associated with program \#3907.
This paper is based on observations collected at the European Organisation for Astronomical Research in the Southern Hemisphere under ESO programme \#108.223D.001.

\vspace{1em}
\noindent\textbf{Funding Statement}

Y.Z. acknowledges the support from the Yangyang Development Fund.
R.F. acknowledges support from the grant PID2023-146295NB-I00, and from the Severo Ochoa grant CEX2021-001131-S funded by MCIN/AEI/10.13039/501100011033 and by ``European Union NextGenerationEU/PRTR’'.
K.T. was supported by JSPS KAKENHI Grant Number JP25K07365 and NAOJ ALMA Scientific Research Grant Code 2025-29B.
M.B. has received funding from the Astronomy \& Astrophysics program of the National Science Foundation (grant No. AST 2510129).
Y.C. was partially supported by Grant-in-Aid for Scientific Research (KAKENHI  number JP24K17103 and 26K00748) of the JSPS.
G.G. gratefully acknowledges support by the ANID BASAL project FB210003.
Z.-Y.L. is supported in part by NSF AST-2308199.

\vspace{1em}
\noindent\textbf{Author Contributions}

Y.W. performed most of the data analysis and drafted the manuscript.
Y.Z. led the project, developed the initial idea, conducted the ALMA and VLA data reduction, and participated in drafting manuscript.
R.F. conducted the JWST and VLT data reduction and analysis, and participated in drafting manuscript.
K.E.I.T. and V.R. contributed to the data acquisition and interpretation, and contributed to the discussions.
K.Y. contributed to data analysis.
J.C.T. contributed to the conceptualization, and participated in drafting manuscript.
M.A., M.T.B, M.B, Y.C, J.M.D.B, Y.D, G.G, P.G, Z.Y.L and Y.L.Y discussed the results and commented on the manuscript.

\vspace{1em}
\noindent\textbf{Competing Interests}

The authors declare no competing interests.

\clearpage

\setcounter{figure}{0}
\renewcommand{\figurename}{Extended Data Fig.}
\renewcommand{\figureautorefname}{Extended Data Fig.}

\renewcommand{\tablename}{Extended Data Table.}


\begin{figure}[ht!]
\centering
\includegraphics[width=1.0\textwidth]{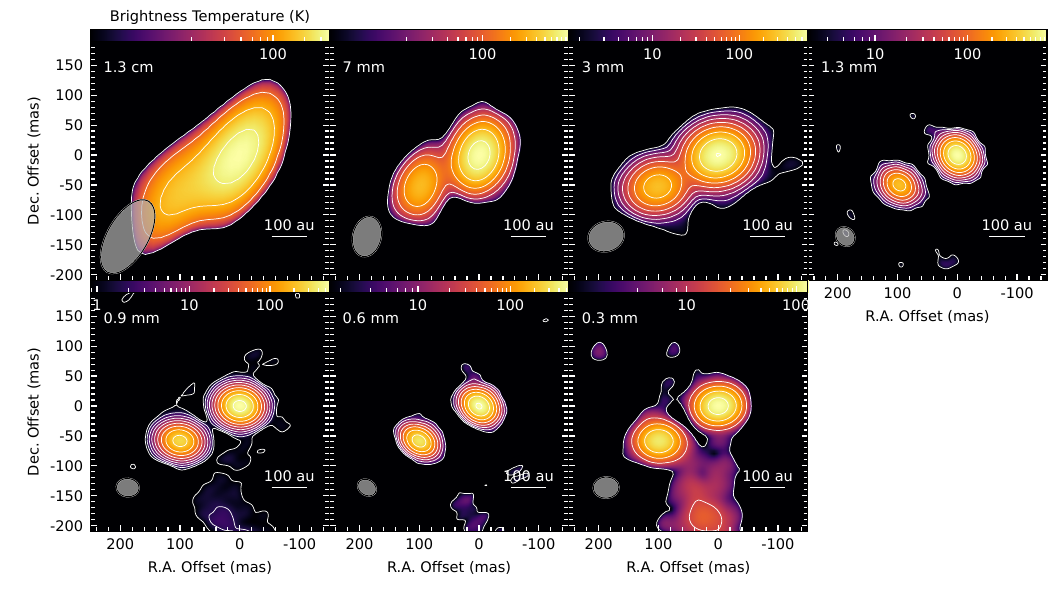}
\caption{
\textbf{Continuum images from 1.3~cm to 0.3~mm.} 
Contours are plotted at $2^{n}\times3\sigma$ ($n=0,1,2,\ldots$), 
where $\sigma$ is the rms noise of each image: 
0.043, 0.034, 0.019, 0.065, 0.034, 0.21 and 2.0 mJy beam$^{-1}$ 
(equivalent to 11, 7.4, 0.91, 1.4, 0.29, 1.3 and 2.1 K, respectively). 
The synthesized beams (lower left) have FWHMs and position angles of 
$137.4\times67.7$~mas ($-30.3^{\circ}$), $67.8\times47.2$~mas ($-13.4^{\circ}$), $60.6\times50.0$~mas ($-73.8^{\circ}$), $34.6\times29.4$~mas ($44.9^{\circ}$), 
$38.1\times30.6$~mas ($84.7^{\circ}$), $32.8\times25.7$~mas ($57.7^{\circ}$) 
and $43.1\times35.6$~mas ($-87.0^{\circ}$), respectively. 
Scale bars indicate physical scale of 100 au in each panel.
Position offsets are measured relative to Source A in each panel.
}\label{extfig:continuum_imgs}
\end{figure}

\clearpage

\begin{figure}[ht!]
\centering
\includegraphics[width=1.0\textwidth]{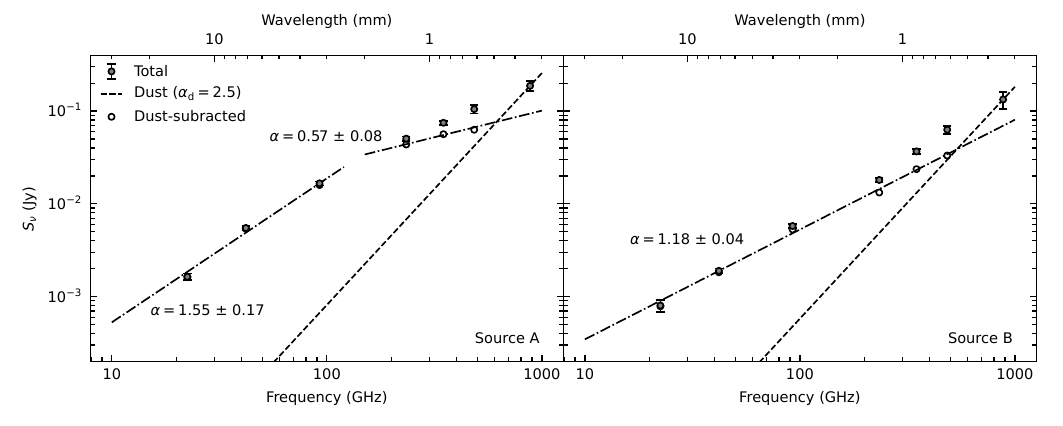}
\caption{
\textbf{Spectral energy distributions from 1.3~cm to 0.3~mm of the binary.} 
Flux densities are from 2D Gaussian fitting (Supplementary Table~1). 
Filled circles denote the total flux densities, 
whereas open circles denote the dust-subtracted flux densities. 
Error bars combine uncertainties from Gaussian fitting, 
flux calibration, and additional systematic contributions 
arising from different array configurations and 
the flux differences between the lower and upper sidebands.
Dashed lines show the dust component assuming $\alpha_{\mathrm{d}}=2.5$ and 
that all 0.3~mm emission is dust. 
Dash–dotted lines show power-law fits to the dust-subtracted emission. 
For Source A, two frequency ranges are fitted separately.
}\label{extfig:sed}
\end{figure}
\clearpage
\begin{figure}[ht!]
\centering
\includegraphics[width=1.0\textwidth]{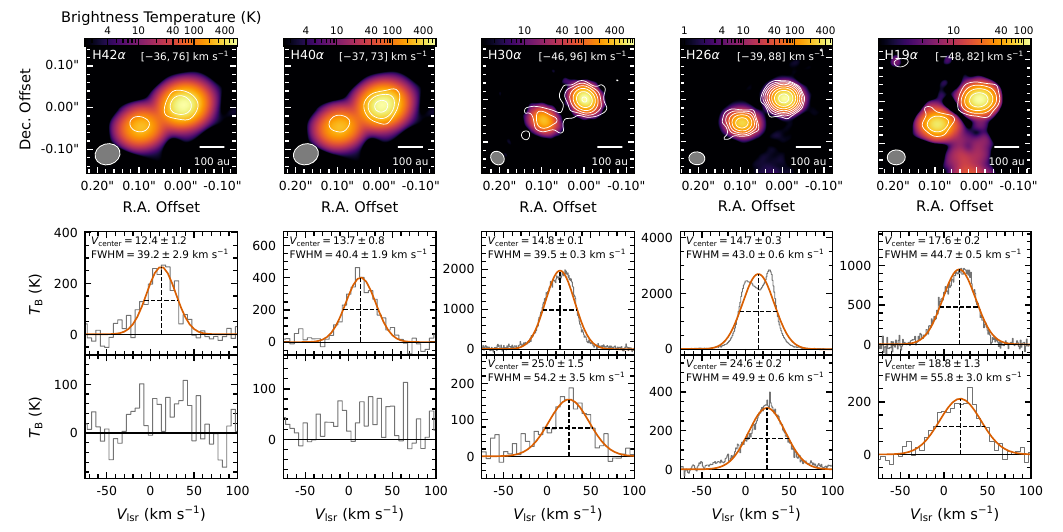}
\caption{
\textbf{Hydrogen recombination line images and spectra.}
{\bf Top:} integrated emission (moment 0) maps of H42$\alpha$, H40$\alpha$, H30$\alpha$, H26$\alpha$, and H19$\alpha$ as contours overlaid on the continuum; 
velocity ranges are shown in each panel. 
Contours are $3\sigma\times2^n$ ($n=0,1,2,\dots$) 
with $\sigma=$ 22, 21, 38, 23, 1239 mJy beam$^{-1}$ km s$^{-1}$.
Scale bars indicate physical scale of 100 au in each panel.
{\bf Middle:} HRL spectra of Source A extracted at the continuum peaks; 
velocity resolutions are $1~\kms$ for strong lines and $5~\kms$ for weaker lines. 
Central velocities and line widths are indicated by dashed lines (see Supplementary Table~2). 
{\bf Bottom:} same for Source B; Gaussian fits are omitted for H42$\alpha$ and H40$\alpha$ due to low S/N.
}
\label{extfig:HRL}
\end{figure}
\clearpage
\begin{figure}[ht!]
\centering
  \includegraphics[width=\textwidth]{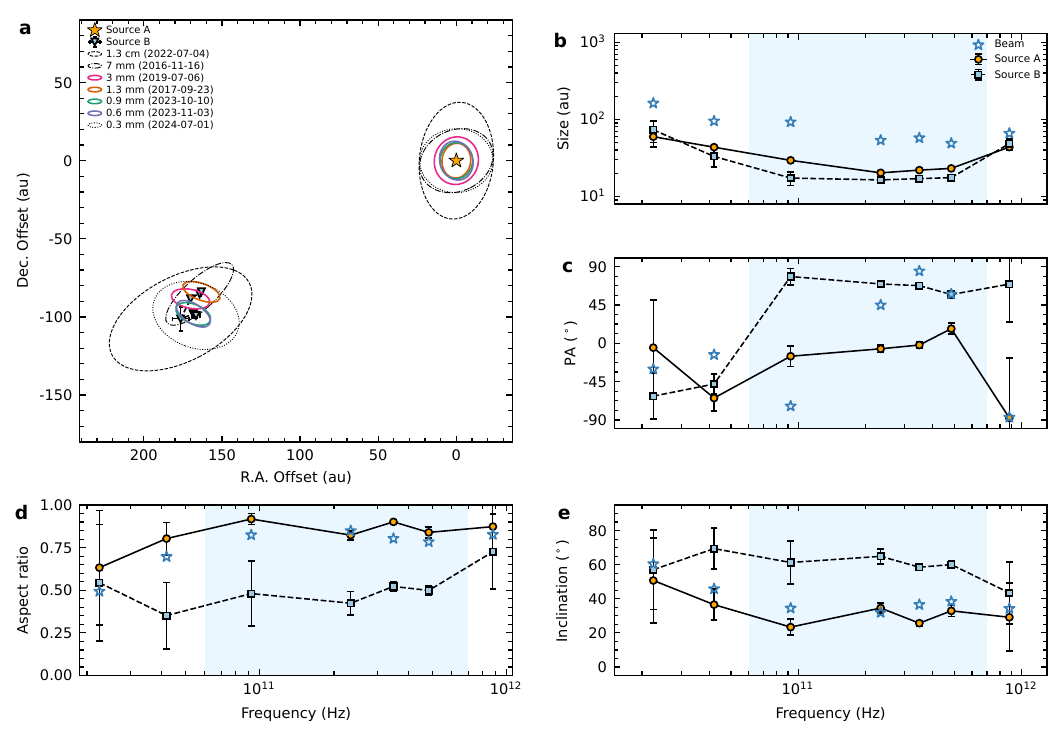}\\
  \caption{\textbf{Deconvolved continuum morphology across observing bands.}
{\bf (a):} Deconvolved 2D Gaussian FWHM profiles of the two sources at different bands shown by ellipses of different colors and line styles (observation epochs labeled in the legends).
The fitted position of sources A and B are marked by the star and triangle symbols, respectively.
{\bf (b):} Deconvolved source sizes as a function of frequency, calculated as $\sqrt{a_\mathrm{decon}\times b_\mathrm{decon}}$, where $a_\mathrm{decon}$ and $b_\mathrm{decon}$ are the deconvolved major and minor axis FWHMs (Supplementary Table~1).
{\bf (c):} Major axis position angles of the deconvolved components versus frequency.
{\bf (d):} Aspect ratios ($b_\mathrm{decon}/a_\mathrm{decon}$) versus frequency.
{\bf (e):} Disk inclinations relative to the plane of the sky, 
estimated from the aspect ratios assuming the deconvolved ellipses trace inclined disks. 
In panels b–e, Source A and B are shown by yellow and blue markers, respectively. 
Error bars denote the uncertainties in deconvolved sizes, position angles, aspect ratios, and inclinations, propagated from the uncertainties of the Gaussian image fitting.
The blue open stars indicate values derived from the synthesized beams for comparison, and shaded regions mark bands with mutually consistent measurements.
}
  \label{extfig:deconvolvedsize}
  \end{figure}

\clearpage

\begin{figure}[h!]
  \begin{center}
  \includegraphics[width=0.7\textwidth]{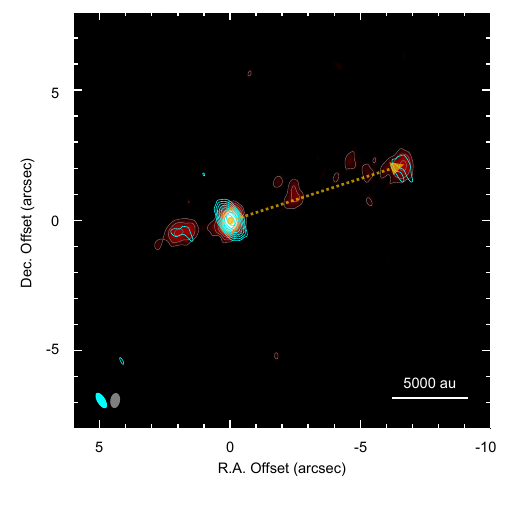}\\
  \caption{\textbf{Continuum images of the central source and radio jets.} The images are observed with VLA 6~cm (background and grey contours) and 1.3~cm B configuration (cyan contours). 
  Jet direction is indicated by the orange arrow. 
  Contours correspond to $4\sigma\times 2^n$ ($n=0,1,\dots$), with $1\sigma=0.0039~\mathrm{mJy~beam^{-1}}$ ($0.63~\mathrm{K}$) for 6~cm and $1\sigma=0.0092~\mathrm{mJy~beam^{-1}}$ ($0.11~\mathrm{K}$) for 1.3~cm.
  The ellipses in the lower-left corner indicate the synthesized beam sizes at the two bands.
  }
  \label{extfig:radiojet}
  \end{center}
\end{figure}

\clearpage


\begin{figure}[ht!]
\centering
  \includegraphics[width=0.9\textwidth]{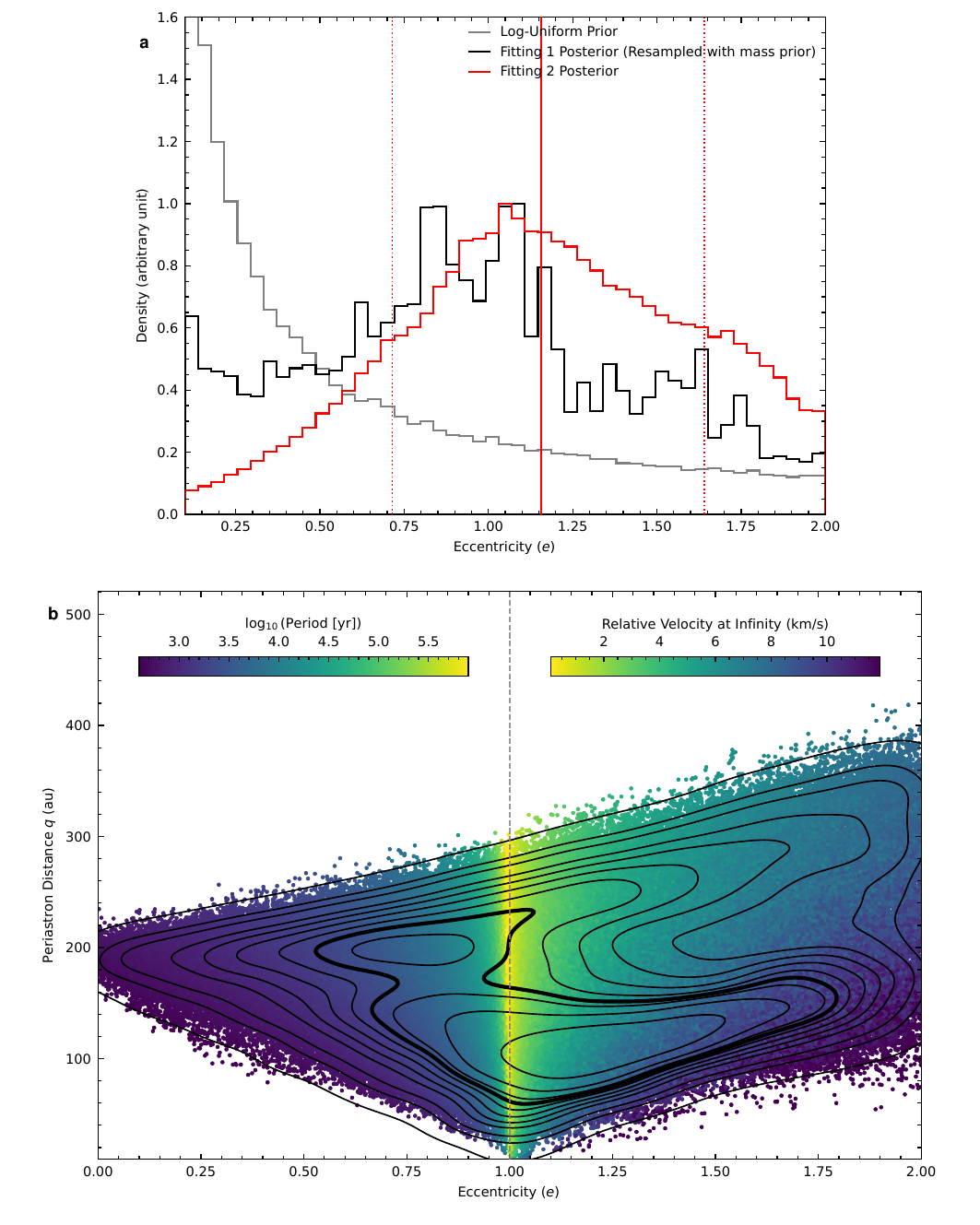}\\
  \caption{\textbf{Eccentricity constraints from the mass-constrained orbital fit.}
    {\bf (a):}
    Comparison of posterior eccentricity distributions from Fitting 1 (black) and Fitting 2 (red). The posterior from Fitting 1 is resampled using a truncated Gaussian prior on $M_\text{tot}$ ($25.9\pm4.0~\msun$ and hard bounds of $17.7$–$28.1~\msun$), consistent with that adopted in Fitting 2. The log-normal eccentricity prior used in Fitting 1 is also shown in grey. The vertical red lines mark the 16th, 50th (median), and 84th percentiles of Fitting 2's posterior distribution.
    {\bf (b):}
    Posterior distribution of periastron distance $q$ as a function of orbital eccentricity $e$ for Fitting 2. 
    Black contours show the kernel density estimate, with the $1\sigma$ credible region emphasized in bold. 
    The vertical grey line at $e=1$ delineates bound (elliptical) and unbound (hyperbolic) regimes. 
    The color scale encodes orbital period for elliptical solutions (left) and velocity at infinity for hyperbolic solutions (right).
  }
  \label{extfig:unified_models_gmassprior}
  \end{figure}


\clearpage


\begin{figure}[ht!]
\centering
  \includegraphics[width=\textwidth]{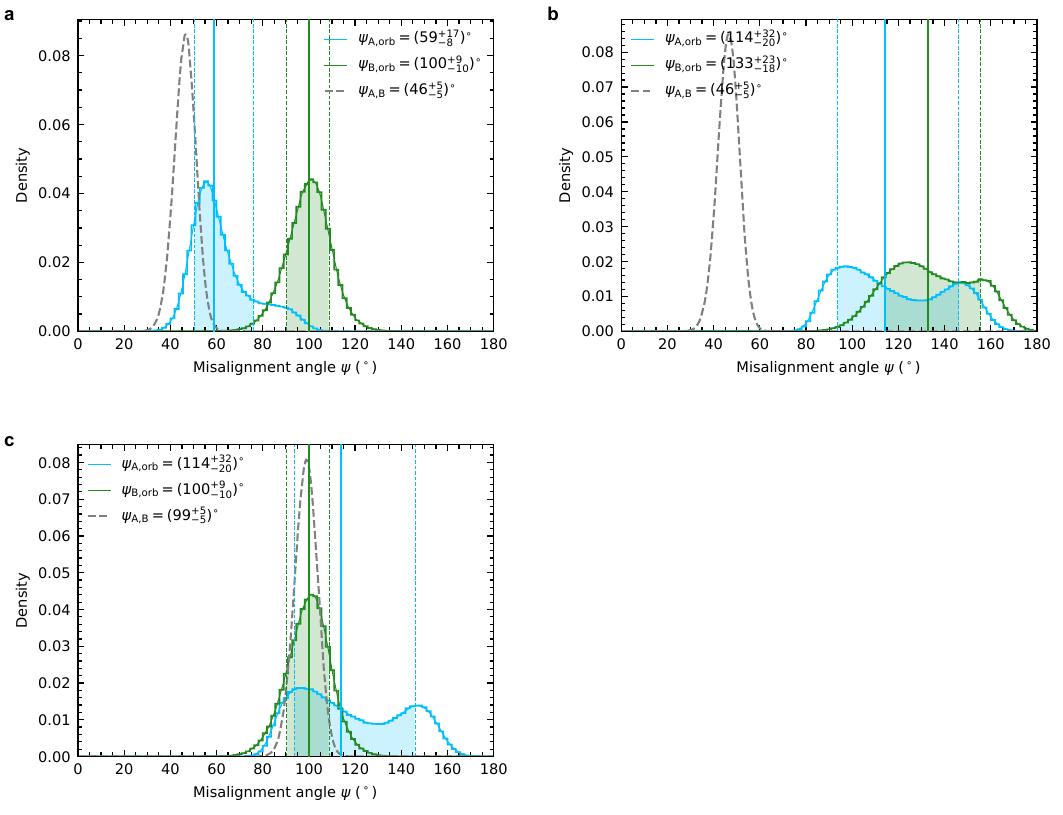}\\
  \caption{\textbf{Disk-orbit misalignments under different disk orientation assumptions.} Distribution of inclination angles between the disk angular momenta and the protobinary orbit are shown for Source A (blue) and Source B (green). The shaded regions indicate the central $68\%$ of the probability distribution, bounded by vertical dashed lines, while the solid lines mark the median. The relative angle between $\mathbf{J}_\text{A}$ and $\mathbf{J}_\text{B}$ is shown with a grey dashed line. {\bf (a)}: Both circumstellar disks have their southeastern halves closer to us.{\bf (b)}: Both circumstellar disks have their northwestern halves closer to us. {\bf (c)}: Source A disk has its northwestern half closer to us while Source B has its southeastern half closer to us.
    }
  \label{extfig:robustness_misalignments}
  \end{figure}

\begin{sidewaystable}
\caption{SED and HRL flux fitting results\label{tab:sed_model_params}}
\begin{center}
\begin{tabular}{cccccccccccc}
\toprule
Source & $\elecm_{0,10\text{ au}}$ & $p$ & $T_e$ & $r_{\text{out}}$ &$i$& $S_{0.3\text{mm},\text{dust}}$ & $\alpha_{\text{d}}$ & $r^*_{0.3\text{mm}}$$^\mathrm{a}$ & $\dot{N_i}$ & $M_{\text{ZAMS}}$ & $M_\mathrm{protostar}$$^\mathrm{b}$ \\
& ($10^{10}~\mathrm{pc} \cdot \mathrm{cm}^{-6}$) & & (K) & (au) & ($^{\circ}$)&(Jy) & & (au) & ($10^{46}$ s$^{-1}$) & ($M_{\odot}$) & ($M_{\odot}$) \\
\midrule
A & $6.7^{+2.2}_{-1.7}$ & $1.9^{+0.3}_{-0.3}$ & $8000$ & $33^{+26}_{-7}$& $53^{+26}_{-35}$ & $0.23^{+0.06}_{-0.05}$ & $2.5$ & $0.7^{+0.4}_{-0.4}$ & $3.6^{+1.8}_{-1.2}$ & $13.4^{+0.5}_{-0.5}$ &8.5-16.3\\
B & $3.1^{+0.6}_{-0.5}$ & $1.9^{+0.3}_{-0.3}$ & $8000$ & $22^{+24}_{-6}$& $60^{+24}_{-39}$ & $0.15^{+0.02}_{-0.02}$ & $2.5$ & $0.5^{+0.3}_{-0.2}$  & $1.6^{+0.8}_{-0.5}$ &$12.5^{+0.5}_{-0.4}$& 7.7-16.5\\
\bottomrule
\end{tabular}
\end{center}
\vspace{0.5ex}
\textbf{Notes.}
(a) Within $r^*_{0.3\mathrm{mm}}$, the optical depth exceeds 3 to be totally optically thick at the lowest wavelength $0.3\mathrm{mm}$. The EM at smaller radii is set be flattened smoothly rather than follow a power-law increase.
(b) Accounting for different accretion histories, the protostellar mass $M_\mathrm{protostar}$ spans a wider range than $M_\mathrm{ZAMS}$.
\end{sidewaystable}

\clearpage

\begin{sidewaystable}
\caption{Summary of orbital fittings\label{tab:orb_comparison}}
\begin{center}
\renewcommand{\arraystretch}{1.3} 
\begin{tabular}{l c c c c}
\toprule
\toprule
\textbf{Orbital Fittings} & \textbf{Fitting 1$^\mathrm{a}$} & \textbf{Fitting 2$^\mathrm{b}$} & \textbf{Fitting 3$^\mathrm{c}$} & \textbf{Fitting 4$^\mathrm{d}$ (Fiducial)} \\
\midrule
Orbital type & Unified & Unified & Elliptical & Parabolic\\
Eccentricity ($e$) prior & Log-Uniform ($0.1-10$) & Uniform ($0-2$) & Uniform ($0-1$) & Fixed at $1.0$\\
Total Mass ($M_{\odot}$) prior & Uniform ($0-100$) & Gaussian ($\mu=25.9, \sigma=4$) & Gaussian ($\mu=25.9, \sigma=4$) & Gaussian ($\mu=25.9, \sigma=4$)\\
Total Mass ($M_{\odot}$) boundary & $0-100$  & $17.7-28.1$ & $17.7-28.1$ & $17.7-28.1$ \\
Spatial Intrinsic scatter (mas) & Fitted & Fitted & Fixed at 0 & Fitted \\
Included Data & Full Epochs & Full Epochs & Single Epoch (1.3 mm) & Full Epochs  \\
\midrule
\multirow{3}{*}{Relevant Figures} & Fig.~\ref{fig:orbit_para}a & Fig.~\ref{fig:misalignment}a,b & \textemdash & Figs.~\ref{fig:orbit_para}b-d and \ref{fig:misalignment}c\\
  & Extended Data Fig.~\ref{extfig:unified_models_gmassprior}a & Extended Data Figs.~\ref{extfig:unified_models_gmassprior} and \ref{extfig:robustness_misalignments}  & \textemdash  & \textemdash  \\
  & Supplementary Fig.~6 & Supplementary Fig.~7 & Supplementary Fig.~8 and 9 & Supplementary Fig.~10\\
\bottomrule
\end{tabular}
\end{center}
\vspace{0.5ex}
\textbf{Notes.} 
(a) A comprehensive exploration of the parameter space.
(b) Fitting with constrained total mass estimated from independent methods.
(c) Fitting based on single-epoch data for comparison with previous study\cite{zhang_dynamics_2019}.
(d) Fiducial parabolic fitting with constrained total mass estimated from independent methods.

\end{sidewaystable}

\clearpage

\begin{table}
\caption{Fiducial parabolic orbital fitting results\label{tab:fiducial_orbit}}
\begin{center}
\begin{tabular}{ccccccc}
\toprule
$q$ & $t_\mathrm{p}$ & $i$ & $\omega$ & $\Omega$ & $\sigma_{\rm int}$ & $s_{\rm 3D}$\\
(au) & ($10^3$ MJD) & ($^{\circ}$) & ($^{\circ}$) & ($^{\circ}$) & (mas) & (au)\\
\midrule
$139^{+88}_{-62}\ (122)$ & $39^{+22}_{-7}\ (38)$ & $53^{+14}_{-7}\ (50)$ & $129^{+20}_{-16}\ (119)$ & $293^{+23}_{-36}\ (291)$ & $1.4_{-0.4}^{+0.6}\ (0.9)$ & $224^{+44}_{-22}$\\
\bottomrule
\end{tabular}
\end{center}
\vspace{0.5ex}
\textbf{Notes.}
Values are listed as posterior medians with $68\%$ credible intervals. Values in parentheses are the corresponding maximum-likelihood values. The three-dimensional separation of the binary
at the current time (2024-07-01) $s_{\rm 3D}$ is also listed.
\end{table}


\clearpage

\setcounter{figure}{0}
\renewcommand{\figurename}{Supplementary Figure}
\renewcommand{\figureautorefname}{Supplementary Figure}

\renewcommand{\tablename}{Supplementary Table}

\noindent{\Large\textbf{Supplementary Information}}

\vspace{1em}

\noindent\textbf{Supplementary Discussion 1. Reliability of continuum image fitting and astrometry measurement}

The deconvolved source morphologies from continuum image fitting are generally consistent from 3 to 0.6 mm (Extended Data Fig.~4). 
However, the deconvolved ellipses at 7 and 13 mm deviate from those measured at higher frequencies, 
likely owing to the larger beam sizes and increased source blending, particularly at 13 mm. 
In addition, the free–free continuum becomes increasingly optically thick toward lower frequencies, 
enlarging the optically thick region and causing departures from the Gaussian assumption. 
The deconvolved morphology at 0.3 mm also differs markedly from that at 3$-$0.6 mm despite a comparable beam size. 
This difference is attributed to the increasing contribution of optically thick dust emission at 0.3 mm, 
which has a spatial distribution distinct from that of the ionized gas traced at longer wavelengths. 
Consistently, the peak brightness temperatures of Sources A and B at 0.3 mm are 133~K and 88~K, respectively, 
indicative of optically thick dust emission. 
Despite these effects, the observed deviations at the higher-frequency bands 
exhibit largely isotropic residuals and are therefore not expected to introduce significant systematic biases 
in the measured source positions. 

The positional uncertainties from Gaussian image fitting generally scale inversely with the peak signal-to-noise ratio. 
However, such uncertainties can underestimate the true astrometric errors 
when systematic effects such as source blending, calibration residuals, 
and departures from the assumed source morphology are present \cite{ellfiterror1997,hernandezgarnica_accurate_2024}. 
In the present data, the astrometric accuracy can be affected by source blending, particularly at 1.3 cm, 
and by intrinsic shifts of the emission peaks arising from wavelength-dependent optical-depth effects. 
To account for such uncharacterized systematics in the proper-motion analysis, 
we introduced an isotropic intrinsic-scatter term, $\sigma_{\rm int}$, 
following the standard approach \cite{Hogg10}. 
The inferred value, $\sigma_{\rm int}=1.3^{+0.5}_{-0.3}$ mas (Fig. 1e,f), 
therefore provides an empirical estimate of the residual astrometric uncertainty not captured by the formal image-fitting errors.

To further assess the robustness of the astrometry, 
we performed visibility-domain fitting in addition to image-domain Gaussian fitting 
(Supplementary Figs.~\ref{extfig:vis_results} and \ref{extfig:vis_vs_imfit}). 
The positional offsets between the two methods are approximately 2 mas at 1.3 cm and approximately 0.5 mas at the other wavelengths. 
These differences are comparable to the inferred intrinsic-scatter term and therefore remain within the adopted astrometric uncertainty budget.

\vspace{1em}
\noindent\textbf{Supplementary Discussion 2. Robustness of the orbital proper-motion measurement}

Several effects could potentially mimic or bias the measured relative motion between the two sources. 
One possibility is wavelength-dependent centroid shifts caused by optical-depth variations. 
If such effects dominated the measured positions, the relative offsets would be expected to follow the ordering of observing wavelength. 
Instead, the measured positional offsets of Source B relative to Source A follow the temporal sequence of the observing epochs 
rather than the wavelength sequence, indicating that the observed shifts are primarily driven by source proper motion.

Another possible origin of the observed position changes is variability associated with radio jets, 
which has been reported in centimeter observations of forming stars\cite{Zapata2015,Cesaroni2023,Cesaroni2024}. 
However, the measured relative motion of approximately 10 km s$^{-1}$ is substantially lower than 
typical jet velocities of 100$-$1000 km s$^{-1}$ \cite{jetreview2018}. 
Moreover, the relative motion remains clearly detectable at millimeter wavelengths, 
where contamination from free–free jet emission is expected to be minimal. 
For example, a positional shift is evident between the 1.3-mm observations obtained in 2017 
and the 0.3-mm observations obtained in 2024. 
Furthermore, although the measured proper-motion direction is broadly aligned with the expected jet axis of Source B, 
no jet is detected toward this source in the free–free continuum (Extended Data Fig.~5).

Taken together, these tests support that the measured relative motion is dominated by binary orbital motion. 
Residual contributions from jet variability, wavelength-dependent optical-depth effects, 
and other systematic uncertainties are incorporated into the intrinsic-scatter term adopted in the astrometric analysis.

\vspace{1em}
\noindent\textbf{Supplementary Discussion 3. Evidence for non-LTE effects in the hydrogen recombination lines}

The observed hydrogen recombination line (HRL) fluxes exhibit a spectral index significantly steeper than 
the value of 1 expected for optically thin LTE emission. 
At low frequencies, such behavior can be partly attributed to increasing free-free continuum optical depth. 
However, the steep trend persists toward higher frequencies, 
where optical-depth effects alone cannot account for the observed fluxes. 
In addition, several high-frequency HRLs display non-Gaussian spectral profiles, 
most notably the H26$\alpha$ line toward Source A (Extended Data Fig.~3).

These characteristics suggest that the HRL emission is affected by departures from LTE. 
One possible explanation is maser amplification, 
which has been predicted and observed for HRLs with principal quantum numbers $n\sim7$--39 \cite{zhang_angular_2017}. 
Such non-LTE effects can modify both the line strengths and spectral profiles, 
potentially contributing to the discrepancies between the observed HRL fluxes and simple LTE expectations. 
Since our SED and HRL fitting adopts an LTE treatment as a first-order approximation,
only the lower-frequency HRL fluxes are fitted (see Method and Fig. 4).

\vspace{1em}
\noindent\textbf{Supplementary Discussion 4. Influence of proper-motion measurements and prior assumptions on orbital constraints}

In previous work on this source\cite{zhang_dynamics_2019}, 
only radial velocity and single-epoch astrometric data were available to constrain the orbital properties. 
To assess the impact of the newly measured proper motions, 
we performed an additional orbital fit (Orbital Fitting 3) using only the same single-epoch 1.3 mm data. 
Although the overall constraint on eccentricity is weak, as expected 
(Supplementary Fig.~\ref{extfig:corner_orbit_ell_single_epoch}), 
the observed relative proper-motion components ($v_{\rm R.A.}$ and $v_{\rm Dec.}$) 
can be reproduced only by models with high eccentricities 
(Supplementary Fig.~\ref{extfig:ell_fit_Band6}). 
This comparison demonstrates that the multi-epoch astrometric measurements provide 
a critical new constraint on the orbital architecture of the system.

In Orbital Fitting 1 and 2, even when the total system mass is constrained by independent observations, 
the eccentricity remains imperfectly determined owing to the limited orbital coverage. 
A pronounced tail toward lower eccentricities is present in the posterior distributions obtained 
from Orbital Fitting 1 and 2 (Fig. 5a, Extended Data Fig. 6a). 
Part of this behavior reflects the adopted prior assumptions. 
In Orbital Fitting 1, a log-uniform prior on eccentricity places greater weight on low-eccentricity solutions, 
resulting in a broader posterior distribution extending toward smaller eccentricities. 
In Orbital Fitting 2, adopting a uniform prior over $0<e<2$ produces a posterior that is more sharply peaked and symmetric around $e\approx1$ 
(see also Supplementary Fig.~\ref{extfig:corner_orbit_unified_fm}). 
Nevertheless, both fitting approaches consistently favor solutions clustered around $e\approx1$ over the independently inferred mass range of the system. 
Lower-eccentricity bound orbits remain viable, but near-parabolic solutions are statistically preferred by the current data.

\clearpage

\begin{figure}[ht!]
\centering
\includegraphics[width=1.0\textwidth]{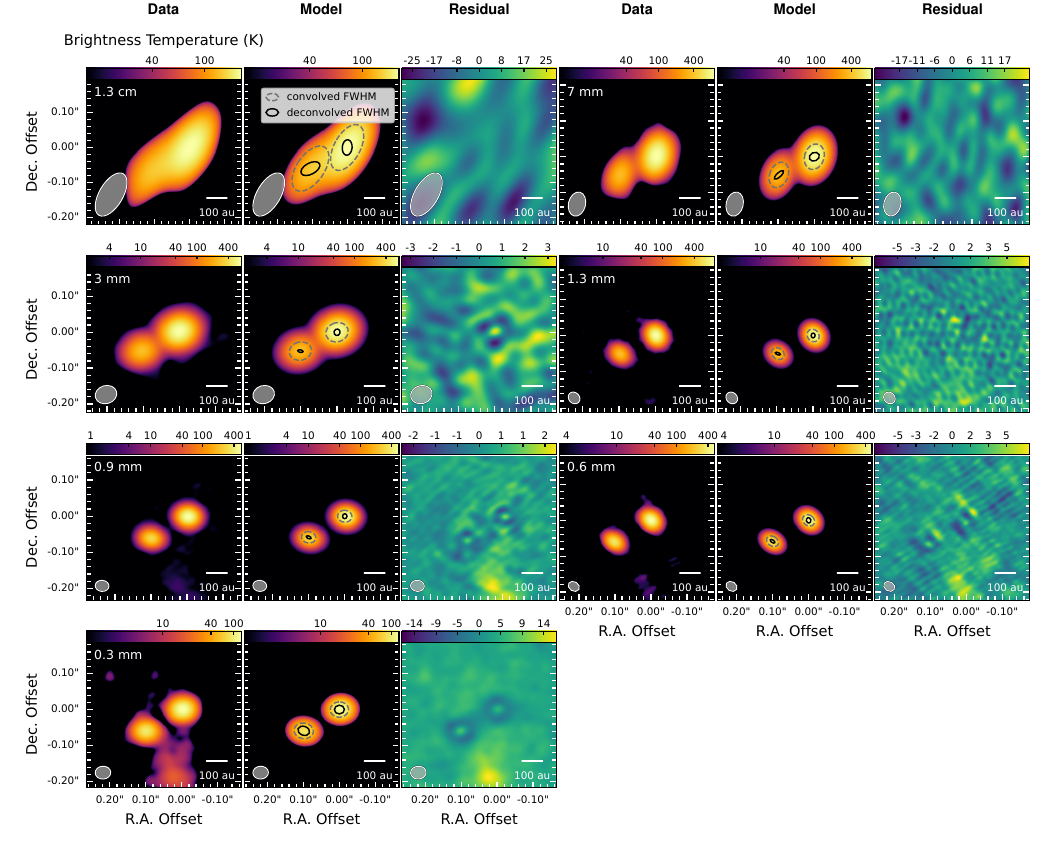}
\caption{
\textbf{2D Gaussian fits to the continuum images.}
In each panel, the convolved and deconvolved Gaussian components are shown 
as grey dashed and black solid ellipses, respectively. 
Grey ellipses at lower-left corners show the synthesized beam sizes.
Scale bars indicate a physical length of 100 au.
Residuals are generally below the $3\sigma$ level.
}\label{extfig:residual}
\end{figure}

\clearpage

\begin{sidewaystable}
\caption{2D Gaussian Fitting to the Continuum Images}
\label{tab:contfit}
\begin{center}
\begin{tabular}{cccccccc}
\toprule
Wavelength & Source & Deconvolved Size (FWHM) & Deconvolved P.A.  & $I_\nu^{\mathrm{peak}}$  & $S_\nu$$^\mathrm{a}$  & $\Delta\alpha$$^\mathrm{b}$ & $\Delta\delta$$^\mathrm{c}$  \\
(mm)&&(mas$\times$mas)&($^{\circ}$)&(K)&(mJy)& (mas) & (mas)\\
\midrule
\multirow{2}{*}{13} & A & $45\pm16~\times~28\pm11$ & $175\pm56$ & $369.7 \pm 9.7$ & $1.64\pm0.13$ & \multirow{2}{*}{$105.02\pm3.13$} & \multirow{2}{*}{$-60.26\pm4.60$} \\
& B & $59\pm26~\times~32\pm15$ & $118\pm27$ & $166.7 \pm 10.0$ & $0.80\pm0.12$ & & \\
\midrule
\multirow{2}{*}{7} & A & $29\pm2~\times~23\pm2$ & $116\pm15$ & $968.5 \pm 7.3$ & $5.5\pm0.3$ & \multirow{2}{*}{$97.79\pm0.46$} & \multirow{2}{*}{$-50.76\pm0.77$} \\
& B & $33\pm5~\times~12\pm6$ & $132\pm12$ & $348.6 \pm 7.2$ & $1.9\pm0.1$ & & \\
\midrule
\multirow{2}{*}{3} & A & $18.3\pm0.4~\times~16.8\pm0.5$ & $165\pm12$ & $710.9 \pm 1.0$ & $16.6\pm0.8$ & \multirow{2}{*}{$101.23\pm0.11$} & \multirow{2}{*}{$-52.63\pm0.08$} \\
& B & $14.9\pm1.4~\times~7.1\pm2.8$ & $78\pm10$ & $261.1 \pm 1.0$ & $5.8\pm0.3$ & & \\
\midrule
\multirow{2}{*}{1.3} & A & $13.3\pm0.2~\times~10.9\pm0.3$ & $173\pm5$ & $965.8 \pm 1.6$ & $50.5\pm2.5$ & \multirow{2}{*}{$97.10\pm0.07$} & \multirow{2}{*}{$-49.97\pm0.07$} \\
& B & $15.0\pm0.6~\times~6.3\pm1.0$ & $70\pm4$ & $355.0 \pm 1.6$ & $18.1\pm0.9$ & & \\
\midrule
\multirow{2}{*}{0.9} & A & $13.76\pm0.08~\times~12.40\pm0.14$ & $178\pm4$ & $560.0 \pm 0.5$ & $75\pm4$ & \multirow{2}{*}{$100.20\pm0.04$} & \multirow{2}{*}{$-58.39\pm0.03$} \\
& B & $14.0\pm0.3~\times~7.3\pm0.3$ & $68\pm4$ & $289.1 \pm 0.5$ & $37\pm2$ & & \\
\midrule
\multirow{2}{*}{0.6} & A & $15.0\pm0.3~\times~12.6\pm0.4$ & $17\pm6$ & $525.4 \pm 1.5$ & $105\pm11$ & \multirow{2}{*}{$100.20\pm0.08$} & \multirow{2}{*}{$-58.32\pm0.07$} \\
& B & $14.8\pm0.4~\times~7.4\pm0.4$ & $57\pm2$ & $340.7 \pm 1.5$ & $63\pm6$ & & \\
\midrule
\multirow{2}{*}{0.3} & A & $28\pm4~\times~24\pm3$ & $92\pm70$ & $133.3 \pm 4.4$ & $188\pm23$ & \multirow{2}{*}{$99.17\pm1.42$} & \multirow{2}{*}{$-58.97\pm1.04$} \\
& B & $34\pm6~\times~25\pm6$ & $69\pm45$ & $88.1 \pm 4.5$ & $133\pm28$ & & \\
\bottomrule
\end{tabular}
\end{center}
\vspace{0.5ex}
\textbf{Notes.}
(a) Uncertainties include contributions from CASA {\it imfit}, flux calibration, and variations between different configurations or LSB/USB images.
(b) Relative R.A. offset of Source B with respect to Source A.
(c) Relative Dec. offset of Source B with respect to Source A.
\end{sidewaystable}
\clearpage

\clearpage

\begin{sidewaystable}
\caption{Properties of HRL emissions\label{tab:HRLs}}
\begin{center}
\begin{tabular}{ccccccccc}
\toprule
HRL & Frequency & Source & Channel Width & Channel Noise& $I_\nu^{\mathrm{peak}}$ & $V_{\text{center}}$& FWHM  & Total Flux$^\mathrm{a}$  \\
&(GHz)&&($\kms$)& ($\mJybeam$, K) & (K) & ($\kms$) &($\kms$) &($\Jykms$) \\
\midrule
\multirow{2}{*}{H42$\alpha$} & \multirow{2}{*}{85.688} & A & 5 & 0.6, 33 & $263\pm26$ & $12.4\pm1.2$ & $39.2\pm2.9$ & $0.22\pm0.01$ \\
 &  & B & 5 & 0.6, 33 & \textemdash & \textemdash & \textemdash & $0.091\pm0.005$ \\
\midrule
\multirow{2}{*}{H40$\alpha$} & \multirow{2}{*}{99.023} & A & 5 & 0.5, 29 & $401\pm25$ & $13.7\pm0.8$ & $40.4\pm2.9$ & $0.37\pm0.02$ \\
 &  & B & 5 & 0.5, 29 & \textemdash & \textemdash & \textemdash & $0.109\pm0.005$ \\
\midrule
\multirow{2}{*}{H30$\alpha$} & \multirow{2}{*}{231.901} & A & 1 & 1.8, 38 & $1965\pm19$ & $14.8\pm0.1$ & $39.5\pm0.3$ & $5.0\pm0.3$ \\
 &  & B & 5 & 0.8, 17 & $155\pm13$ & $25.0\pm1.5$ & $54.2\pm3.5$ & $0.95\pm0.05$ \\
\midrule
\multirow{2}{*}{H26$\alpha$} & \multirow{2}{*}{353.623} & A & 1 & 0.9, 9 & $2697\pm49$ & $14.7\pm0.3$ & $43.0\pm0.6$ & $13.7\pm0.7$ \\
 &  & B & 1 & 0.9, 9 & $318\pm5$ & $24.6\pm0.2$ & $49.9\pm0.6$ & $2.0\pm0.1$ \\
\midrule
\multirow{2}{*}{H19$\alpha$} & \multirow{2}{*}{888.047} & A & 1 & 45.3, 40 & $951\pm15$ & $17.6\pm0.2$ & $44.7\pm0.5$ & $55.8\pm2.8$ \\
 &  & B & 5 & 22.8, 20 & $213\pm15$ & $18.8\pm1.3$ & $55.8\pm3.0$ & $19.4\pm1.0$ \\
\bottomrule
\end{tabular}
\end{center}
\vspace{0.5ex}
\textbf{Notes.}
(a) HRL total flux uncertainties combine the flux calibration error and the photometric error from the moment 0 maps.
\end{sidewaystable}

\clearpage
\begin{figure}[ht!]
\centering
  \includegraphics[width=\textwidth]{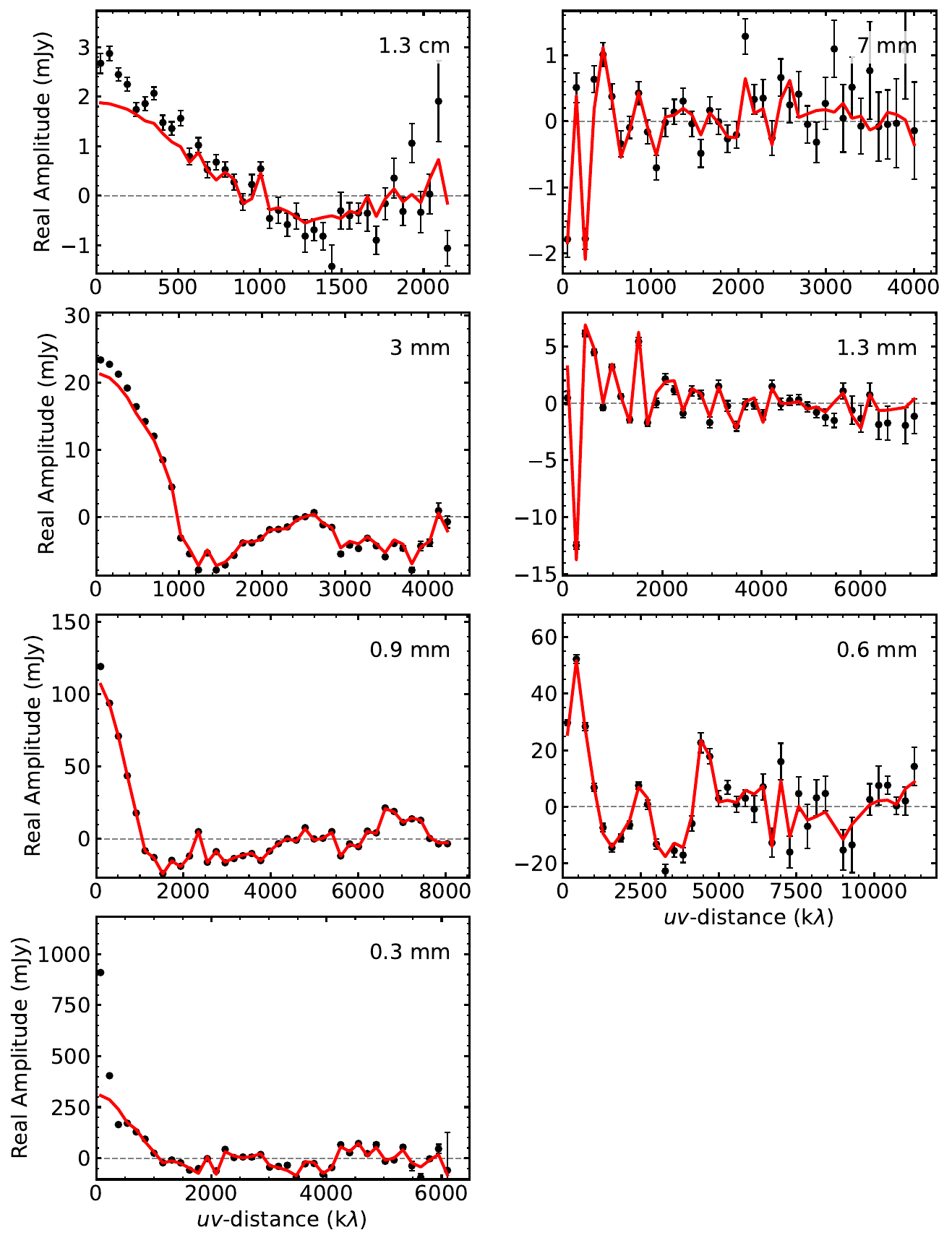}\\
  \caption{
  \textbf{Two-component Gaussian fits to the continuum visibility data.}
  Black points show the real part of the visibilities, binned into 40 points for clarity. Error bars denote the standard error of the mean real visibility in each bin.
  The grey dashed horizontal line marks zero real visibility.
  Red curves show the corresponding best-fitting models.}
  \label{extfig:vis_results}
  \end{figure}

\clearpage

\begin{figure}[ht!]
\centering
  \includegraphics[width=0.85\textwidth]{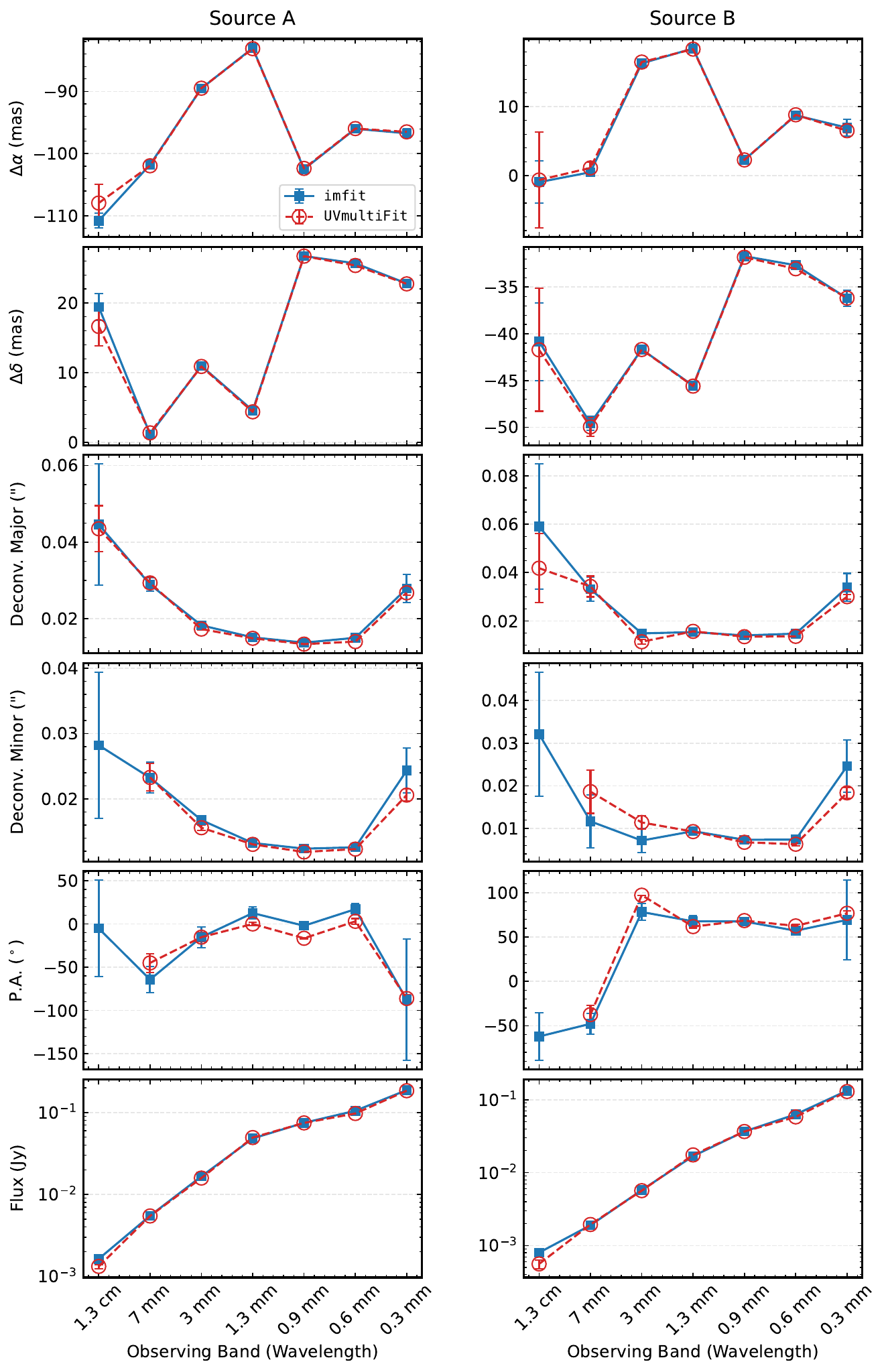}\\
  \caption{
  \textbf{Comparison of continuum parameters derived from visibility fitting and image fitting.} Results from visibility fitting via {\it UVmultiFit} are shown with open red circles while those from image fitting via {\it imfit} are shown with filled blue squares. Error bars indicate uncertainties either derived directly from the fits or propagated from the fitted parameters for the two approaches.
  Shown parameters include absolute source positions, Gaussian properties, and total flux densities of the two sources. 
  Source positions are expressed as offsets relative 
  to $\left(\alpha_{\rm ICRS}, \delta_{\rm ICRS}\right) = (07^{\rm h}32^{\rm m}09.79^{\rm s}, -16^\circ58'12''.15)$. 
  In the K band, the Gaussian aspect ratio was fixed during visibility fitting; 
  consequently, the minor-axis FWHM and position angle are not reported for this band.}
  \label{extfig:vis_vs_imfit}
  \end{figure}

\begin{figure}[ht!]
\centering
  \includegraphics[width=\textwidth]{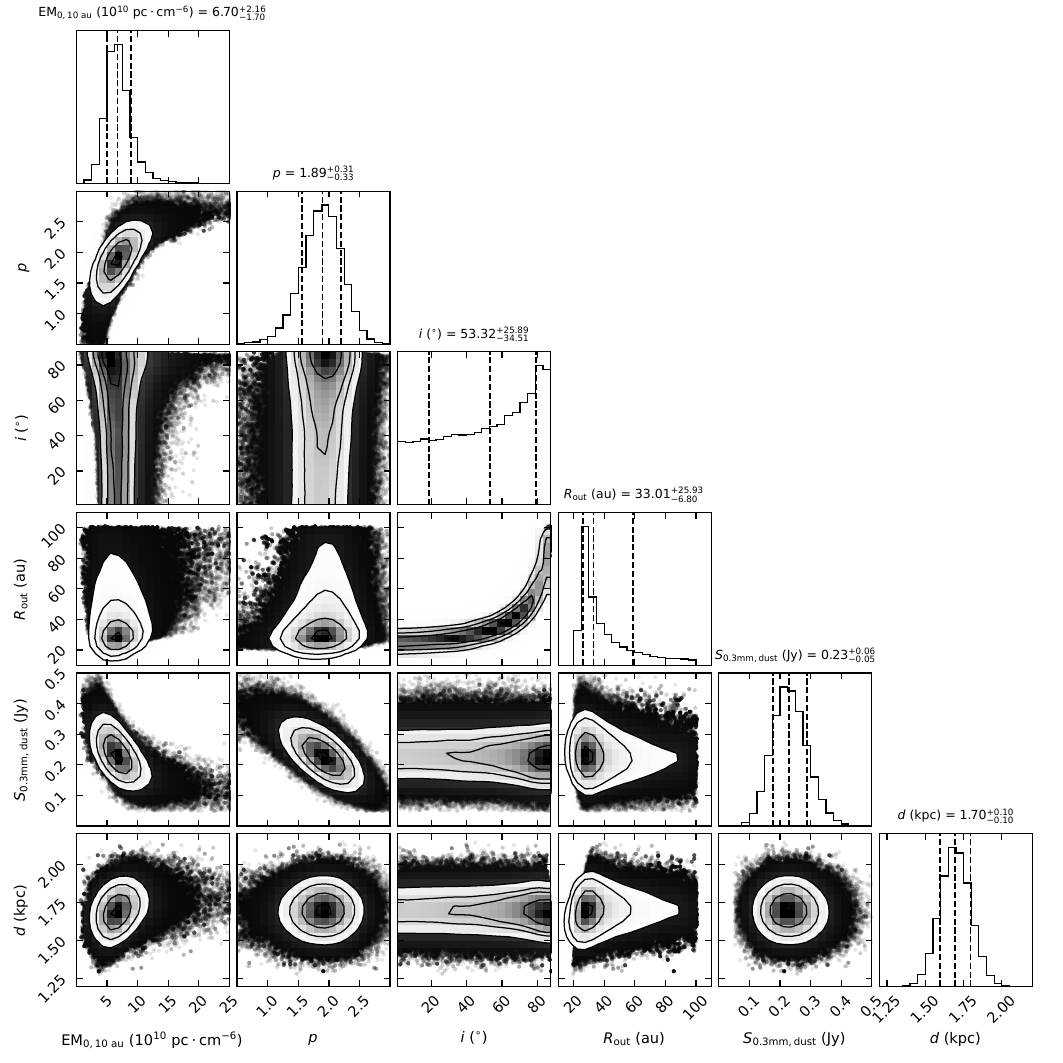}\\
  \caption{\textbf{Posterior distributions from the joint fit to continuum SED and HRL fluxes for Source A.} 
  The corner plot illustrates the distributions and covariances of the six free parameters in the physical model: the emission-measure normalization, $\mathrm{EM}_0$, at the reference radius $r_0 = 10~\mathrm{au}$; the power-law index of the emission-measure distribution, $p$; the outer truncation radius of the ionized disk, $r_{\mathrm{out}}$; the disk inclination, $i$; the dust-continuum normalization at 0.3 mm, $S_{0.3,\mathrm{mm},\mathrm{dust}}$; and the distance to the system, $d$.
  The source distance is treated as a free parameter to account for its uncertainty.
  In the marginal distributions, the dashed lines indicate the 16th, 50th (median), and 84th percentiles.}
  \label{extfig:corner_sed_sA}
  \end{figure}

\clearpage

\begin{figure}[ht!]
\centering
  \includegraphics[width=\textwidth]{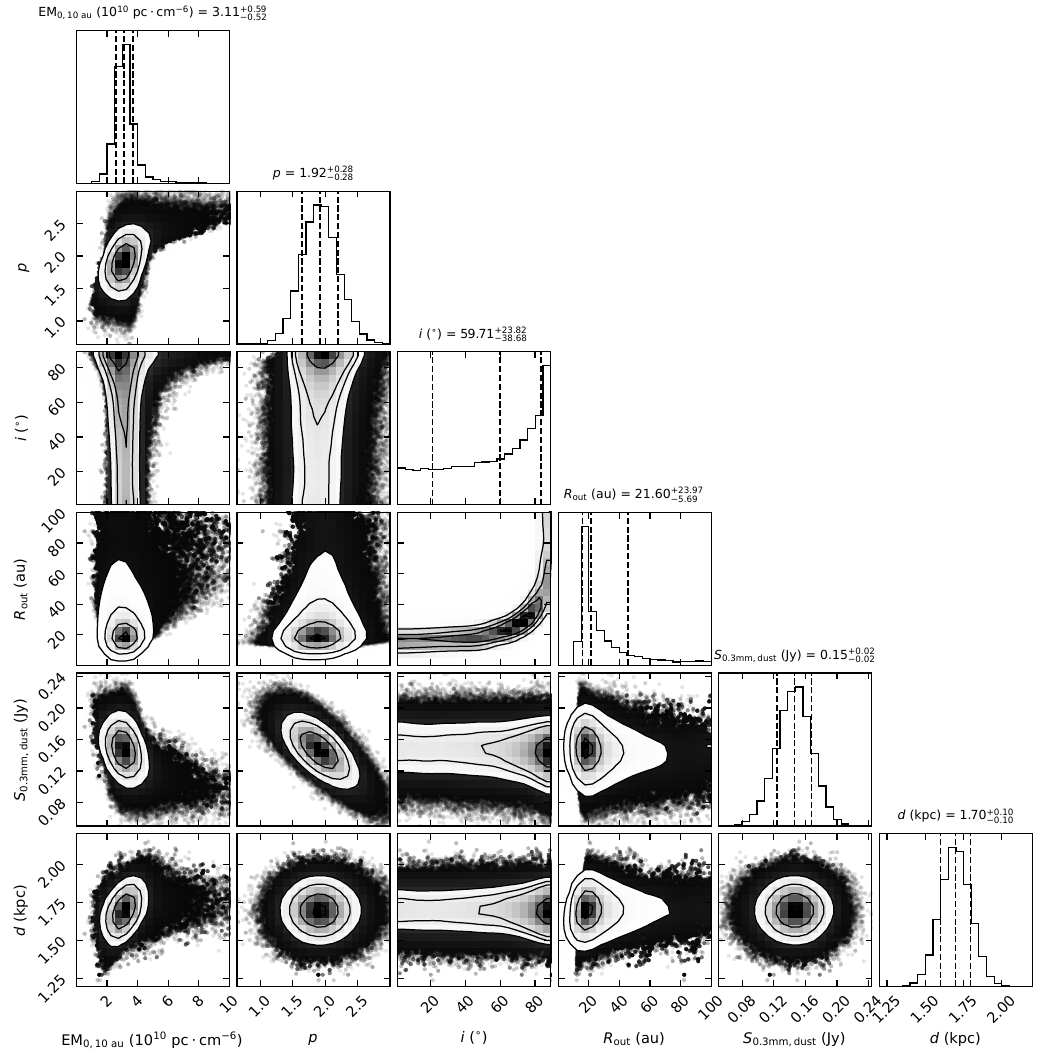}\\
  \caption{\textbf{Posterior distributions from the joint fit to continuum SED and HRL fluxes for Source B.} 
 The corner plot illustrates the distributions and covariances of the six free parameters in the physical model: the emission-measure normalization, $\mathrm{EM}_0$, at the reference radius $r_0 = 10~\mathrm{au}$; the power-law index of the emission-measure distribution, $p$; the outer truncation radius of the ionized disk, $r_{\mathrm{out}}$; the disk inclination, $i$; the dust-continuum normalization at 0.3 mm, $S_{0.3,\mathrm{mm},\mathrm{dust}}$; and the distance to the system, $d$.
  The source distance is treated as a free parameter to account for its uncertainty.
  In the marginal distributions, the dashed lines indicate the 16th, 50th (median), and 84th percentiles.}
  \label{extfig:corner_sed_sB}
  \end{figure}

\clearpage

\begin{figure}[ht!]
\centering
  \includegraphics[width=\textwidth]{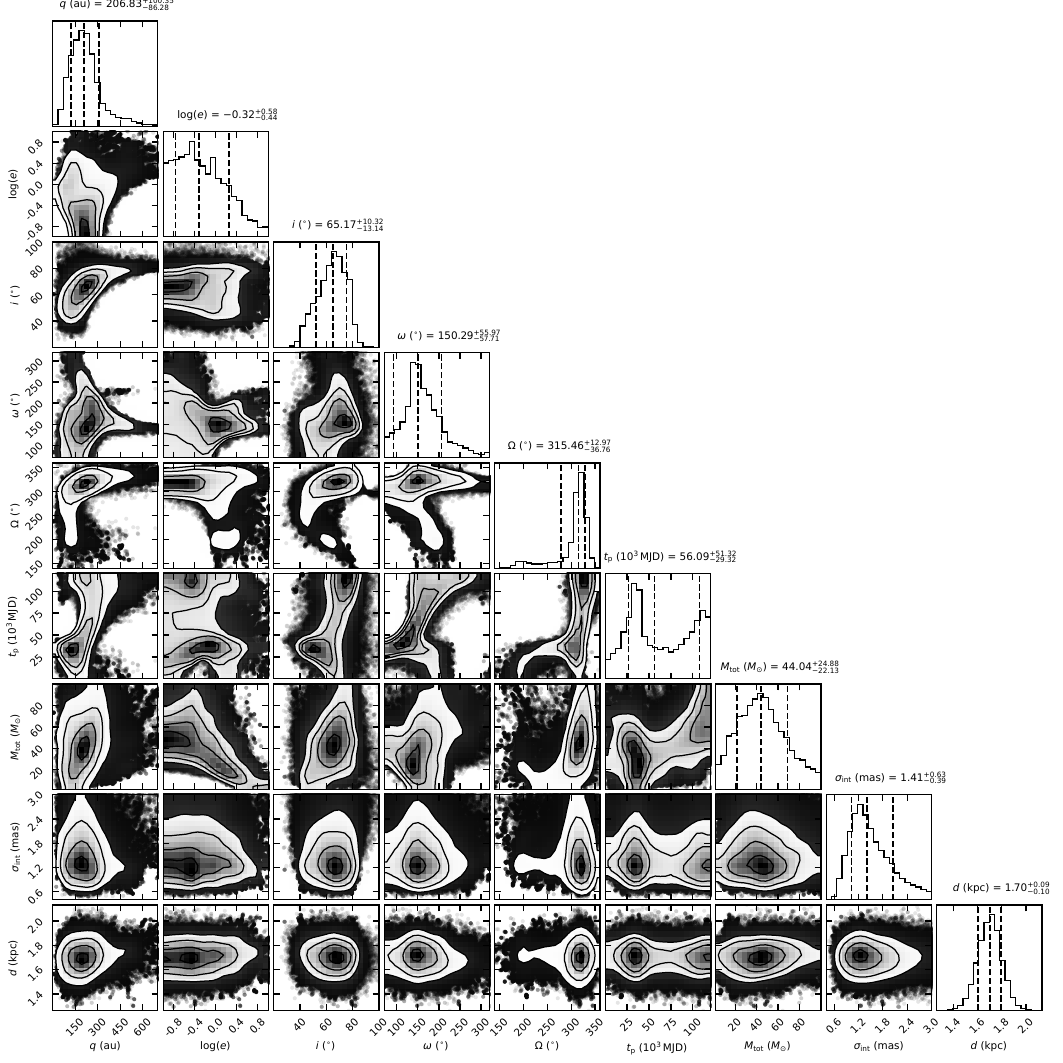}\\
  \caption{\textbf{Posterior samples from the unified orbital fitting (Orbital Fitting 1).} 
  Nine free parameters for the fitting are shown:
  periapsis distance $q$, eccentricity $e$, inclination $i$, argument of periapsis $\omega$, longitude of ascending node $\Omega$, time of periapsis $t_\text{p}$, and total mass $M_\mathrm{tot}$, intrinsic scatter $\sigma_\mathrm{int}$, 
  and distance to this system $d$. The source distance is treated as a free parameter to account for its uncertainty.
  In the marginal distributions, the dashed lines indicate the 16th, 50th (median), and 84th percentiles.}
  \label{extfig:corner_orbit_unified}
  \end{figure}

\clearpage

\begin{figure}[ht!]
\centering
  \includegraphics[width=\textwidth]{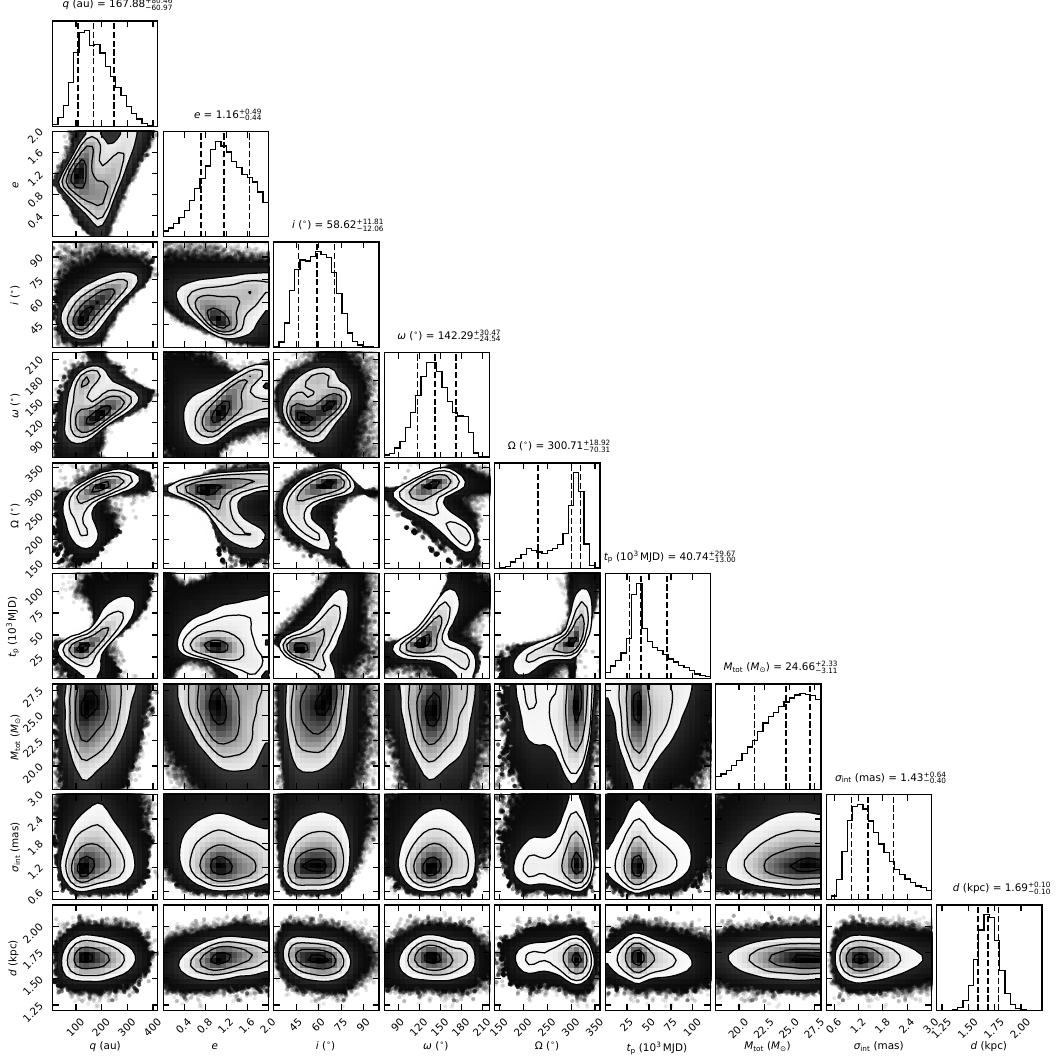}\\
  \caption{\textbf{Posterior samples from the unified orbital fitting with a constrained mass range  (Orbital Fitting 2).} 
  A truncated Gaussian prior for $M_\text{tot}$ is adopted, with mean $25.9~\msun$, standard deviation $4.0~\msun$, and hard bounds of $17.7$–$28.1~\msun$.
  The eccentricity prior is set to be uniform between $0-2$. 
  Nine free parameters are shown: periapsis distance $q$, eccentricity $e$, inclination $i$, argument of periapsis $\omega$, longitude of ascending node $\Omega$, time of periapsis $t_\text{p}$, and total mass $M_\mathrm{tot}$, intrinsic scatter $\sigma_\mathrm{int}$, 
  and distance to this system $d$. The source distance is treated as a free parameter to account for its uncertainty.
  In the marginal distributions, the dashed lines indicate the 16th, 50th (median), and 84th percentiles.
  The fitting has a weighted reduced chi-square of $\chi^2_{\mathrm{reduced}}=1.57^{+0.74}_{-0.54}$ over the full posterior.}
  \label{extfig:corner_orbit_unified_fm}
  \end{figure}

\clearpage

\begin{figure}[ht!]
\centering
  \includegraphics[width=\textwidth]{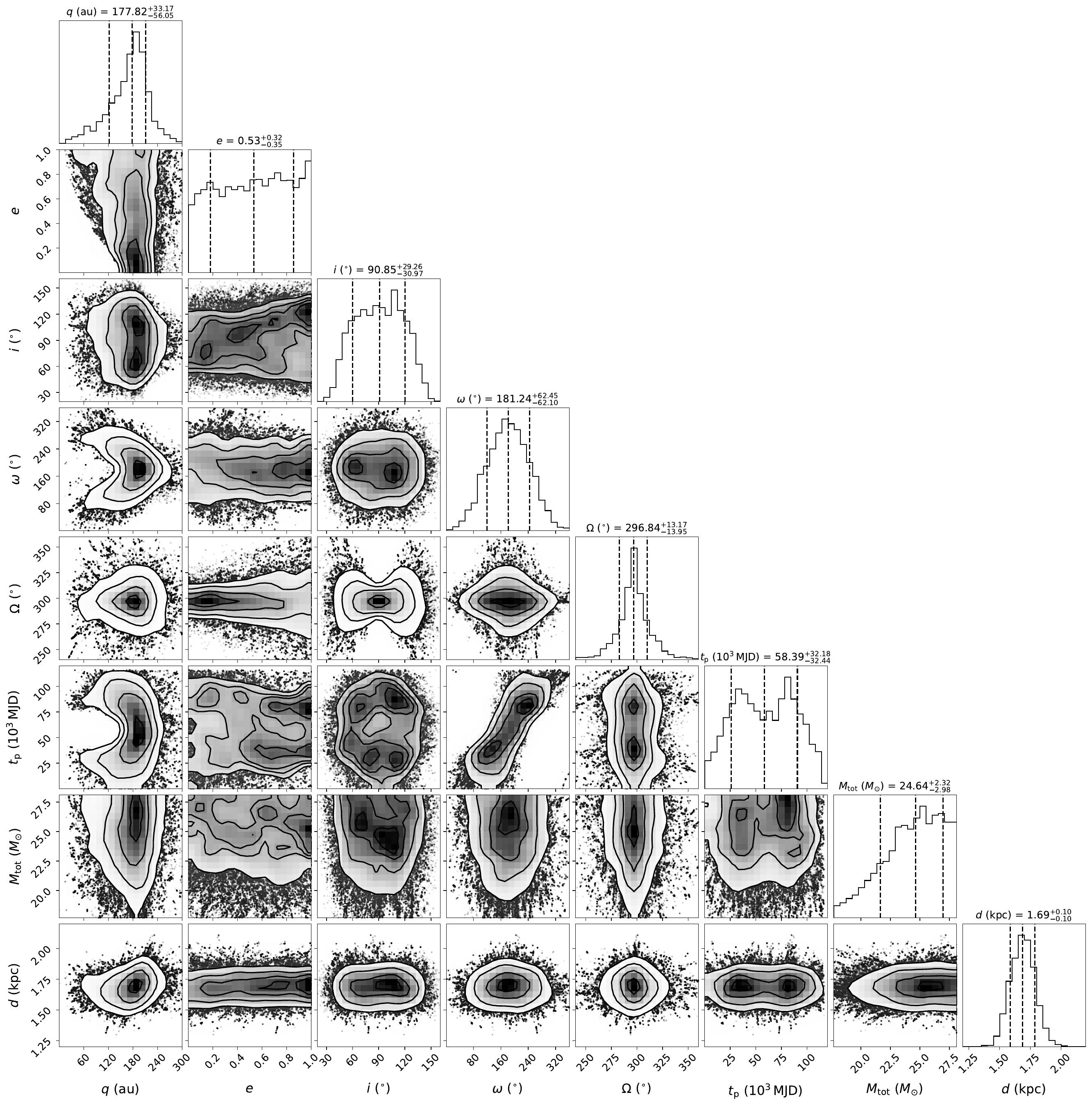}\\
  \caption{
  \textbf{Posterior samples from the elliptical orbital fitting to the 1.3~mm data alone  (Orbital Fitting 3).} 
  The prior of eccentricity is set to uniform between $0-1$ and a truncated Gaussian prior is adopted for $M_\text{tot}$, with mean $25.9~\msun$, standard deviation $4.0~\msun$, and hard bounds of $17.7$–$28.1~\msun$. Eight free parameters are shown:
  periapsis distance $q$, eccentricity $e$, inclination $i$, argument of periapsis $\omega$, longitude of ascending node $\Omega$, time of periapsis $t_\text{p}$, and total mass $M_\mathrm{tot}$ and distance to this system $d$. The source distance is treated as a free parameter to account for its uncertainty.
  In the marginal distributions, the dashed lines indicate the 16th, 50th (median), and 84th percentiles.
  }
  \label{extfig:corner_orbit_ell_single_epoch}
  \end{figure}

\clearpage

\begin{figure}[ht!]
\centering
  \includegraphics[width=0.9\textwidth]{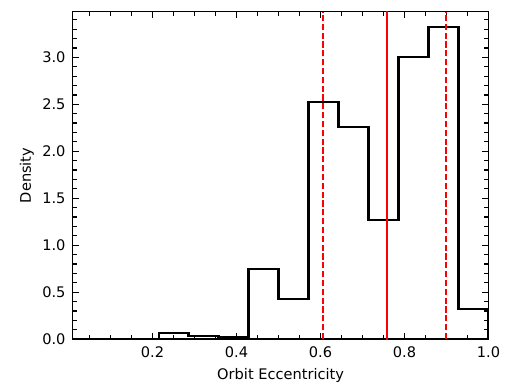}
  \caption{\textbf{Eccentricity constraints from single-epoch orbital fitting (Orbital Fitting 3).}
The eccentricity distribution is derived from the orbit solutions in Fitting 3 that be able to reproduce the observed $v_{\text{R.A.}}$ and $v_{\text{Dec.}}$.
    The 16th, 50th (median), and 84th percentiles are indicated with vertical red lines.}
  \label{extfig:ell_fit_Band6}
  \end{figure}

\clearpage

\begin{figure}[ht!]
\centering
  \includegraphics[width=\textwidth]{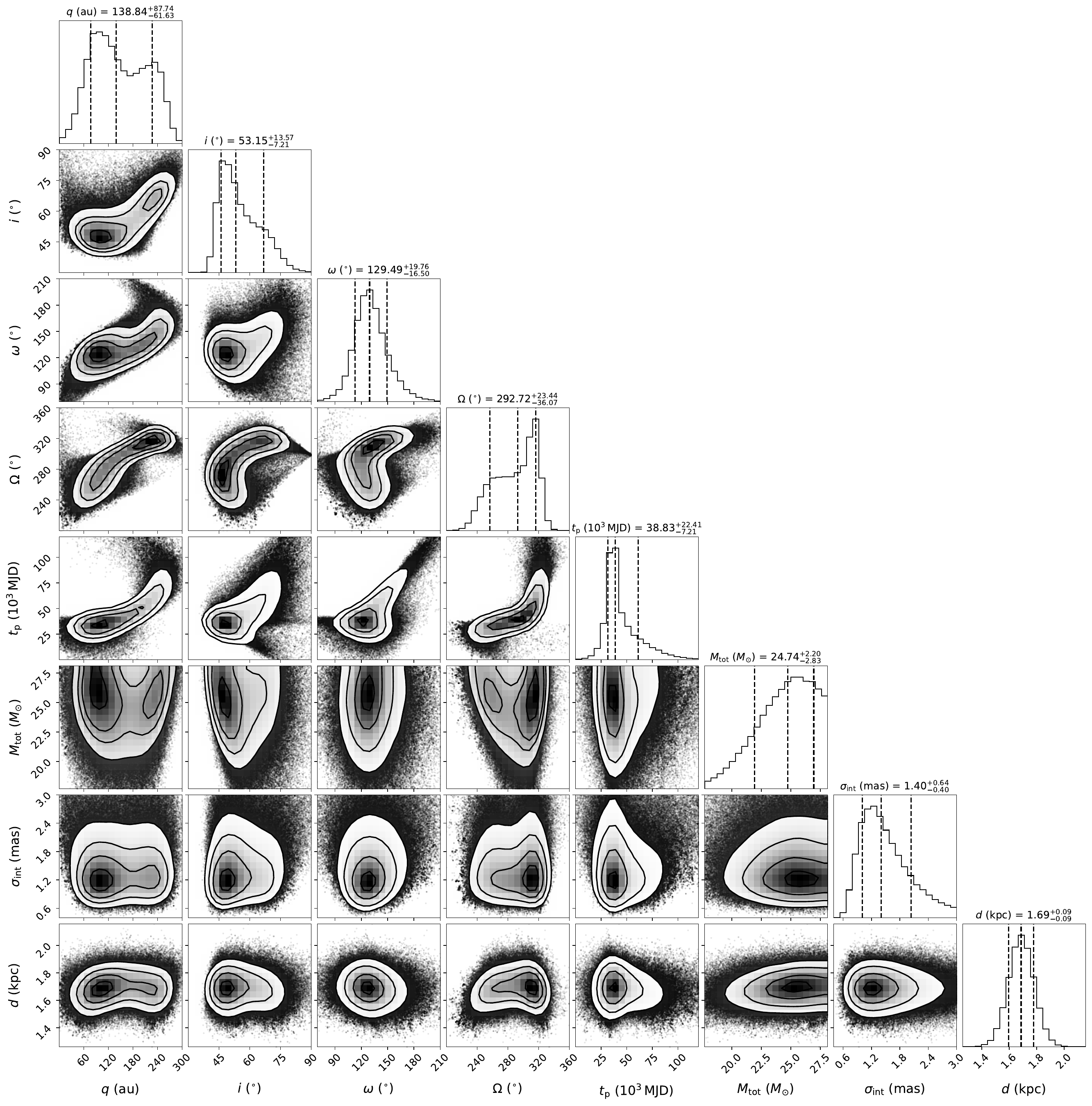}\\
  \caption{\textbf{Posterior samples from the fiducial parabolic orbital fitting with constrained mass range  (Orbital Fitting 4).}
  A truncated Gaussian prior for $M_\text{tot}$ is adopted, with mean $25.9~\msun$, standard deviation $4.0~\msun$, and hard bounds of $17.7$–$28.1~\msun$.
  Eight free parameters are shown:
  periapsis distance $q$, inclination $i$, argument of periapsis $\omega$, longitude of ascending node $\Omega$, time of periapsis $t_\text{p}$, and total mass $M_\mathrm{tot}$, intrinsic scatter $\sigma_\mathrm{int}$, 
  and distance to this system $d$. The source distance is treated as a free parameter to account for its uncertainty.
  In the marginal distributions, the dashed lines indicate the 16th, 50th (median), and 84th percentiles. 
  The fitting has a weighted reduced chi-square of $\chi^2_{\mathrm{reduced}}=1.35_{-0.46}^{+0.65}$ over the full posterior.}
  \label{extfig:corner_orbit_para}
  \end{figure}

\clearpage


\clearpage





\clearpage


\end{document}